\documentclass[journal,12pt,onecolumn,draftclsnofoot,]{IEEEtran}

\usepackage{algorithm}
\usepackage{algpseudocode}
\usepackage{graphicx}
\usepackage{subcaption}
\usepackage{booktabs}
\usepackage{color}

\usepackage{amssymb}
\usepackage{amsmath}

\usepackage{cite}

\begin{document}

\title{Hyperspectral image unmixing with LiDAR~data-aided spatial regularization}

\author{Tatsumi~Uezato, Mathieu Fauvel and Nicolas Dobigeon
\thanks{Part of this work has been funded by EU FP7 through the ERANETMED
JC-WATER program, MapInvPlnt Project ANR-15-NMED-
0002-02 and by the MUESLI IDEX ATS  project, Toulouse INP.}
\thanks{T. Uezato and N. Dobigeon are with University of Toulouse, IRIT/INP-ENSEEIHT, CNRS, 2 rue Charles Camichel, BP 7122, 31071 Toulouse Cedex 7, France (e-mail: \{Tatsumi.Uezato, Nicolas.Dobigeon\}@enseeiht.fr).}
\thanks{M. Fauvel is with University of Toulouse, INRA, DYNAFOR, BP 32607, Auzeville-Tolosane, 31326 Castanet Tolosan, France (e-mail: mathieu.fauvel@ensat.fr).}
        }% <-this % stops a space

\maketitle

\begin{abstract}
Spectral unmixing methods incorporating spatial regularizations have demonstrated increasing interest. Although  spatial regularizers which promote smoothness of the abundance maps have been widely used, they may overly smooth these maps and, in particular, may not preserve edges present in the hyperspectral image. Existing unmixing methods usually ignore these edge structures or use edge information derived from the hyperspectral image itself. However, this information may be affected by large amounts of noise or variations in illumination, leading to erroneous spatial information incorporated into the unmixing procedure. This paper proposes a simple, yet powerful, spectral unmixing framework which incorporates external data (i.e. LiDAR data). The LiDAR measurements can be easily exploited to adjust standard spatial regularizations applied to the unmixing process. The proposed framework is rigorously evaluated using two simulated datasets and a real hyperspectral image. It is compared with competing methods that rely on spatial information derived from a hyperspectral image. The results show that the proposed framework can provide better abundance estimates and, more specifically, can significantly improve the abundance estimates for pixels affected by shadows.
\end{abstract}


\newpage
\section{Introduction}

\IEEEPARstart{S}{pectral} unmixing (SU) has been used for a wide variety of applications~\cite{Kesha2002}. SU consists in decomposing spectral mixtures at the sub-pixel scale and estimates quantitative abundances of materials~\cite{Biouc2012}. In order to estimate accurate abundances, a variety of spectral unmixing methods have been developed~\cite{Biouc2012,Dobig2009,Thouv2016a}. Among them, methods that incorporate spatial information have proven to be a valuable approach~\cite{Shi2014,Uezat2016a,Iorda2012}. A class of such methods assumes that a hyperspectral image is composed of multiple spatially homogeneous regions where abundances of materials share same statistical moments~\cite{Eches2011,Eches2013}. The methods first classifies the hyperspectral images into multiple homogeneous regions and then promote similar abundance estimates within each segmented region. Another group of methods assumes that two neighboring pixels have similar abundances and show smooth transitions in abundances~\cite{Uezat2016a,Iorda2012,Bauer2014,Drume2016}. These methods promote similar abundance estimates in a local neighborhood.\\
Recently, SU methods that promote the spatial homogeneity in a local neighborhood have received an increasing attention~\cite{Bauer2014,Iorda2012,Uezat2016a}. These methods usually relies on $\ell_2$-norm \cite{Thouv2016,Zhou2016}, $\ell_1$-norm or total variation (TV) \cite{Iorda2012,Uezat2016a,Jie2014} regularizations to describe the spatial variations of the abundance  maps in a local neighborhood. The use of the $\ell_2$-norm of the abundance map gradient generally overly smooths edges and may not be suitable for estimating abundances when considering a hyperspectral image with significant structured patterns~\cite{Bauer2015}. Conversely, the TV-based regularization is known to better preserve edges in the abundance maps. However, it may also introduce shrinkage effects, in particular for pixels located on edges between areas characterized by similar yet different abundance maps.\\
While the edge-preserving property is important and has been actively studied in hyperspectral image classification~\cite{Kang2014,Xia2016,Tarab2010}, there are a few studies that consider edge-preserving spectral unmixing. Edges are usually located in areas of junction of distinct materials and cause abrupt changes in abundance maps. Once localized, these spatial changes can be incorporated into the unmixing process by weighting the spatial regularization \cite{Liu2012,Castr2011,Zhong2014}.

To encode this spatial information, the existing methods commonly use a so-called \emph{guidance} map, e.g., derived from the hyperspectral image or from the abundance maps. However, these strategies exclusively relying on the hyperspectral image  to be unmixed or associated derived products may not be suitable. Indeed, the hyperspectral image may be greatly affected by variations in illumination~\cite{Uezat2016,Murph2012,Uezat2016b}, leading to unreliable weighing function. Another issue results from the fact that some materials (e.g., road or roof) show similar spectral shapes~\cite{Ni2014}. In this case, by computing the weights directly from the hyperspectral image, some structures can be missed. In both cases, these weights cannot correctly describe the correlation in the local neighborhood pixels, which may significantly impact the relevance of the resulting weighted spatial regularization.\\
From these findings, it clearly appears that a guidance image must be robust to the aforementioned problems. LiDAR data have great potential to provide meaningful spatial information regarding scenes where the elevation differences play an important role for discriminating different objects. For example, LiDAR data enable the edges between  spectrally similar vegetation classes to be identified thanks to their different heights~\cite{Ke2010}. More generally, LiDAR data have been successfully used for classification of hyperspectral images and show improvement in classification accuracy~\cite{Ke2010,Dalpo2008,Ni2014}. Conversely, only a few studies have used LiDAR data for spectral unmixing. However, such data can be useful to design an appropriate guidance map, i.e., a weighting function to be incorporated into the spatial regularizations. In~\cite{Castr2012}, the authors used LiDAR data to calculate the weights describing spatial correlations in a local neighborhood for constraining spatial regularization. This study showed that including the weights into the spatial regularization can improve abundance estimates for regions that are partially occluded by a shadow. However, this study did not investigate whether LiDAR data can be combined with other guidance maps derived from the hyperspectral image or abundance maps. This paper proposes to fill this gap.

More precisely, the contributions of this paper are twofold: 1) it develops a general spectral unmixing framework which allows external data (i.e., LiDAR data) to be easily incorporated to calculate weights and guide the spatial regularization; and 2) it provides a comprehensive comparison of the weighting functions derived from the LiDAR data, the hyperspectral image or abundance maps. The various instances of the proposed unmixing framework have been rigorously validated using simulated data and hyperspectral imagery. It aims at evaluating whether the LiDAR data used for spatial regularization can improve significantly the estimates of abundances.\\

\makeatletter
\newcommand{\rmnum}[1]{\romannumeral #1}
\newcommand{\Rmnum}[1]{\expandafter\@slowromancap\romannumeral #1@}
\makeatother
The paper is organized as follows. Section \Rmnum{2} introduces the proposed spectral unmixing framework which allows a conventional spatial regularization to be adjusted according to guidance maps derived from internal (e.g., hyperspectral data, abundance maps) or external (e.g., LiDAR) data. Experiments using simulated and real data are reported in Section \Rmnum{3} and \Rmnum{4}, respectively. Finally, the paper is concluded in Section \Rmnum{5}.

\section{Proposed method}
\subsection{LiDAR data-aided spatial regularization}
The linear mixture model (LMM) has been widely used to decompose a mixed spectrum into a collection of ``pure'' spectra known as \textit{endmembers} and their abundances. LMM represents a mixed spectrum as the linear combination
\begin{eqnarray}
  \mathbf{y}_i=\mathbf{E}\mathbf{a}_i+\mathbf{n}_i.
\end{eqnarray}
where $\mathbf{y}_i \in \mathbb{R}^{L \times 1}$ is the mixed $L$-spectrum measured in the $i$th pixel, $\mathbf{E} \in \mathbb{R}^{L \times M}$ is the matrix of $M$ endmembers, $\mathbf{a}_i \in \mathbb{R}^{M \times 1}$ represents the abundance fractions in the same spatial location, $\mathbf{n}_i \in \mathbb{R}^{L \times 1}$ is an additive term associated with modeling error and noise measurements.
%represents noise in the $i$ th pixel, $L$ is the number of bands and $M$ is the number of endmembers.
The abundance non-negativity constraint (ANC) and the abundance sum-to-one constraint (ASC) are usually imposed as follows
\begin{eqnarray}
\label{eq:ASC_ANC}
  \forall i,\  \forall m, \ a_{mi}\geq 0,\quad \text{and} \quad \forall i \ \sum_{m=1}^{M}a_{mi}=1.
\end{eqnarray}
where $a_{mi}$ is the abundance fraction of the $m$th endmember in the $i$th pixel with $\mathbf{a}_i=\left[a_{1i},\ldots,a_{Mi}\right]^T$. The endmember signatures $\mathbf{e}_1,\ldots,\mathbf{e}_M$ can be chosen from a given spectral library or extracted from the hyperspectral image. Once the endmember matrix $\mathbf{E}$ has been fixed, spectral unmixing reduces to the estimation of the abundance vectors $\mathbf{a}_i$ ($i=1,\ldots,N$, where $N$ is the number of pixels), which can be formulated as the following optimization problem
\begin{equation}
	 \operatornamewithlimits{min}_{\mathbf{a}_i}\frac{1}{2}\Vert\mathbf{y}_i - \mathbf{E}\mathbf{a}_i\Vert^2_2, \  \text{subject to } \eqref{eq:ASC_ANC}.
\end{equation}
In the optimization problem introduced above, abundance vectors $\mathbf{a}_i$ are estimated for each pixel independently, ignoring the possible spatial information underlying the hyperspectral image. In order to take advantage of the spatial information, estimation of these abundance vectors should be conducted jointly, where the optimization criterion is complemented by a spatial regularization ${\phi(\cdot)}$, i.e.,
\begin{equation}
\label{eq:opti_problem}
	 \operatornamewithlimits{min}_{\mathbf{A}}\frac{1}{2}\Vert \mathbf{Y} - \mathbf{E} \mathbf{A}\Vert^2_2 + \lambda \phi(\mathbf{A}), \  \text{subject to } \eqref{eq:ASC_ANC}.
\end{equation}
where $\mathbf{Y} = \left[\mathbf{y}_1,\ldots,\mathbf{y}_N\right]$ and $\mathbf{A} = \left[\mathbf{a}_1,\ldots,\mathbf{a}_N\right]$ are the matrices of measurements and abundances, respectively, and $\lambda$ is the regularization parameter which controls the balance between the data fitting term and the spatial regularization. Various regularizations have been considered in the literature to promote spatial coherence of the abundance maps, see for instance \cite{Eches2011,Eches2013,Iorda2012,Bauer2014,Drume2016,Uezat2016a}. One popular choice consists in resorting to the $\ell_p$-norm of the finite differences, i.e.,
\begin{equation}
\label{noweight}
 \phi(\mathbf{A}) = \sum_{i=1}^N \sum_{j \in \mathcal{N}(i)}\Vert\mathbf{a}_i-\mathbf{a}_j\Vert_p^p.
\end{equation}
where $\mathcal{N}(i)$ is the set of pixels in the neighborhood\footnote{In this work, the conventional 4-order neighborhood structure will be considered for simplicity.} of the $i$th pixel. In particular, when $p=2$, smooth transitions in the abundance maps are expected. Conversely, the specific case $p=1$ leads to the anisotropic total variation (TV) penalization, which is known to promote piecewise homogeneous abundance maps \cite{Jie2014}.

It is worth noting that, in \eqref{noweight}, each neighboring pixel equally contributes to the spatial regularization term. However, this property may be inappropriate for pixels located in edges. Indeed, neighboring pixels belonging to different objects are expected to contribute to the spatial regularization differently. To alleviate this issue, it is natural to weight this spatial regularization, as in ~\cite{Liu2012,Zhong2014,Feng2015},
\begin{equation}
 \phi(\mathbf{A}) = \sum_{i=1}^N \sum_{j \in \mathcal{N}(i)} w_{ij} \Vert\mathbf{a}_i-\mathbf{a}_j\Vert_p^p.
 \label{eq:w_phi}
\end{equation}
where $w_{ij}$ is a weight describing the spatial similarity between $\mathbf{a}_i$ and $\mathbf{a}_j$. In particular, when the $i$th and $j$th pixels correspond to two distinct objects in the image, their respective abundance vectors $\mathbf{a}_i$ and $\mathbf{a}_j$ are likely to be dissimilar and the weight $w_{ij}$ can be tuned to zero. The set of weights $\left\{w_{ij}\right\}_{ij}$ can be estimated using a so-called guidance map, which gathers relevant spatial and edge information. It can be computed directly from the hyperspectral image, or learned from the abundance maps  \cite{Liu2012,Zhong2014,Feng2015}. Another strategy, widely adopted when conducting hyperspectral image classification, consists in computing the first principal component (PC) of the hyperspectral image~\cite{Kang2014}. However, deriving the guidance maps from the hyperspectral data itself, i.e., by considering the hyperspectral measurements or associated quantities such as abundance maps or principal components, suffer from several drawbacks. First, the hyperspectral image can be corrupted by measurement noise or illumination variations, which would significantly impact the relevance and confidence of the resulting computed weights. Moreover, the information present in the hyperspectral image may not be sufficient to properly identify distinct yet neighboring objects, resulting in inappropriate spatially coherent regularization. In such cases, the digital surface model (DSM) derived from LiDAR data represents a promising and interesting opportunity to guide the guidance map elaboration. Indeed, since DSM encodes the object heights, its variation is expected to be highly correlated with the spatial distribution of the endmembers, implicitly identifying edges between non-homogeneous regions. Thus, as highlighted earlier, in particular the the edge areas, DSM can be useful to \emph{switch off} the spatial regularization by imposing $w_{ij}=0$ for the couples of pixels $(i,j)$ located in these edges. Besides, one great advantage of the LiDAR-based DSM is that it is not impacted by possible illumination conditions. This property is particularly useful within urban or forest areas where the height of objects plays an important role~\cite{Ni2014,Dalpo2008}. This paper proposes to capitalize on the availability of LiDAR as complementary data to incorporate the DSM model into the unmixing process. Moreover, it shows that this complementary guidance maps can be easily coupled with other more conventional guidance maps derived from the hyperspectral data, e.g., in terms of measurements, principal components or abundances.

\subsection{Different types of weights}\label{section_weight}

A variety of guidance images can be used to adjust the weights $w_{ij}$ in the spatial regularization \eqref{eq:w_phi}. In what follows, in addition to the DSM-based guidance map, three distinct guidance maps are presented, as well as their combinations with the DSM one. These weighting functions are introduced in the following paragraphs.

\subsubsection{Weights derived from the hyperspectral image (w-HI)}
The most natural way to derive weights consists in estimating the similarity between pixels within the hyperspectral image itself, i.e.,
\begin{eqnarray}
\label{eq:weight_HI}
	w_{ij}=\frac{1}{Q_i}\exp \left(-\frac{1}{\sigma_y^2}\frac{\Vert \mathbf{y}_i - \mathbf{y}_j \Vert_2^2}{\Vert \mathbf{y}_i+ \mathbf{y}_j\Vert_2^2} \right).
\end{eqnarray}
where $\mathbf{y}_i$ and $\mathbf{y}_j$ are the spectra measured in the $i$th and $j$th pixels of the hyperspectral image, $\sigma_y^2$ is a parameter controlling the weight range and $Q_i$ is a normalizing constant ensuring $\sum_{j \in \mathcal{N}(i)}w_{ij}=1$. The resulting weighting function will be referred to as \emph{w-HI}.

\subsubsection{Weights derived from principal components (w-PC$K$)}
\label{subsubsec:weights}
Principal component analysis (PCA) is known to concentrate most of the useful information into a few components. Formally, it transforms the $L\times N$-data matrix
$\mathbf{Y}$ into the $K\times N$ matrix $\mathbf{P}$ of $K$ principal components, with $K \leq L$. Weights can be estimated from the similarity between pixels of the principal components
\begin{eqnarray}
\label{eq:weight_PC}
	w_{ij}=\frac{1}{Q_i} \exp \left( -\frac{1}{\sigma_p^2}\frac{\left\| \mathbf{p}_i - \mathbf{p}_j \right\|^2}{\left\|\mathbf{p}_i + \mathbf{p}_j\right\|^2} \right).
\end{eqnarray}
where $\mathbf{p}_i$ and $\mathbf{p}_j$ are the $i$th and $j$th pixels of the principal component matrix $\mathbf{P}$, $\sigma_p^2$ is the parameter adjusting the weight range. In what follows, only the first principal component will be considered, i.e., $K=1$ and the resulting weighting function will be referred to as \emph{w-PC1}.

\subsubsection{Weights derived from abundances (w-A)}
The similarity between pixels can be computed from the abundance maps, leading to the following guidance map denoted \emph{w-A}
\begin{equation}
\label{eq:weight_A}
	w_{ij}=\frac{1}{Q_i}\exp\left(-\frac{1}{\sigma_a^2}\frac{\Vert \mathbf{a}_i - \mathbf{a}_j \Vert_2^2}{\Vert \mathbf{a}_i+ \mathbf{a}_j\Vert_2^2}\right).
\end{equation}
where $\mathbf{a}_i$ and $\mathbf{a}_j$ are the abundance vectors associated with the the $i$th and $j$th pixels, $\sigma_a^2$ is a parameter controlling the weight range. Note that, within an unmixing framework, this guidance map cannot be computed directly since relying on unknown abundance quantities.

\subsubsection{Weights derived from the digital surface model (w-DSM)}
When LiDAR provides DSM as complementary data, the weights can be adjusted by computing the similarity between neighboring pixels from their respective heights
\begin{equation}
\label{eq:weight_DSM}
	w_{ij}=\frac{1}{Q_i}\exp\left(-\frac{1}{\sigma_h^2}\frac{( h_i - h_j )^2}{(h_i + h_j)^2}\right).
\end{equation}
where $h_i$ and $h_j$ are the heights associated with the $i$th and $j$th pixels provided by DSM and $\sigma_h^2$ is a parameter controlling the weight range.

% \subsubsection{Weights derived from PC1 and DSM (w-PC1-DSM)}
\subsubsection{Combining DSM and other guidance maps}
\label{subsubsec:combining_DSM}
Based on the previous definitions, the guidance maps can be derived by combining the
similarity between neighboring pixels estimated by DSM and other quantities, such as the hyperspectral image, leading to the \emph{w-HI-DSM} weighting function
\begin{equation}
\label{eq:weight_HI_DSM}
	\begin{split}
	 w_{ij}=\frac{1}{Q_i} &\left[ \exp\left(-\frac{1}{\sigma_y^2}\frac{\Vert \mathbf{y}_i - \mathbf{y}_j \Vert_2^2}{\Vert \mathbf{y}_i+ \mathbf{y}_j\Vert_2^2}\right) \right. \\ & \left. \qquad + \exp\left(-\frac{1}{\sigma_h^2}\frac{( h_i - h_j )^2}{(h_i + h_j)^2}\right) \right].
    \end{split}
\end{equation}
the first principal component, leading to the \emph{w-PC1-DSM} weighting function
\begin{equation}
\label{eq:weight_PC_DSM}
	\begin{split}
	 w_{ij}= \frac{1}{Q_i} &\left[ \exp\left(-\frac{1}{\sigma_p^2}\frac{( p_i - p_j )^2}{(p_i + p_j)^2}\right) \right. \\ & \left. \qquad  + \exp\left(-\frac{1}{\sigma_h^2}\frac{( h_i - h_j )^2}{(h_i + h_j)^2}\right) \right].
    \end{split}
\end{equation}
or the abundances, leading to the \emph{w-A-DSM} weighting function
\begin{equation}
\label{eq:weight_A_DSM}
	\begin{split}
	 w_{ij}=\frac{1}{Q_i} & \left[ \exp\left(-\frac{1}{\sigma_a^2}\frac{\Vert \mathbf{a}_i - \mathbf{a}_j \Vert_2^2}{\Vert \mathbf{a}_i+ \mathbf{a}_j\Vert_2^2}\right) \right. \\ & \left. \qquad + \exp\left( -\frac{1}{\sigma_h^2}\frac{( h_i - h_j )^2}{(h_i + h_j)^2}\right) \right].
    \end{split}
\end{equation}

\subsection{Optimization}
This paragraph details the optimization procedure implemented to solve \eqref{eq:opti_problem} when the regularization term is defined by \eqref{eq:w_phi} with the weighting functions introduced in Section \ref{section_weight}. More precisely, in this work, we consider the case where the  function $\phi(\cdot)$ in \eqref{eq:w_phi} is chosen as a TV-regularization, i.e., with $p=1$. By denoting $\mathbf{W}=\left[ \mathbf{W}_\leftarrow \mathbf{W}_\rightarrow \mathbf{W}_\uparrow \mathbf{W}_\downarrow \right] \in \mathbb{R}^{N \times 4N}$ the sparse matrix associated with the weighted difference operator between a target pixel and the neighboring pixels in the four canonical directions, estimating the $N$ abundance vectors can be rewritten as the following optimization problem
\begin{equation}
	\min\limits_{\mathbf{A}}\frac{1}{2}\Vert\mathbf{Y} - \mathbf{E}\mathbf{A}\Vert^2_F +\lambda\Vert\mathbf{AW}\Vert_{1,1} +\iota_{\mathcal{N}}(\mathbf{A}) +\iota_{\mathcal{S}}(\mathbf{A}).
\end{equation}
where $\iota_{\mathcal{C}}(\cdot)$ is the indicator function of the set $\mathcal{C}$ defined by
\begin{equation}
 \iota_{\mathcal{C}}(\mathbf{u}) = \left\{
  \begin{array}{ll}
    1, & \hbox{if $\mathbf{u} \in \mathcal{C}$;} \\
    \infty, & \hbox{otherwise.}
  \end{array}
\right.
\end{equation}
The convex sets $\mathcal{N}$ and $\mathcal{S}$ are associated with the nonnegativity and additivity constraints \eqref{eq:ASC_ANC} defined by
\begin{eqnarray}
 \mathcal{N}& = \left\{\mathbf{U}=\left[\mathbf{u}_1,\ldots,\mathbf{u}_N\right] \in\mathbb{R}^{M\times N}: \mathbf{u}_i  \succeq \boldsymbol{0}_M\right\}.\\
 \mathcal{S}& = \left\{\mathbf{U}=\left[\mathbf{u}_1,\ldots,\mathbf{u}_N\right] \in\mathbb{R}^{M\times N}: \boldsymbol{1}_M^T \mathbf{u}_i  =1\right\}.
\end{eqnarray}
where $\boldsymbol{0}_M$ and $\boldsymbol{1}_M$ are the $M$-dimensional vectors composed of $0$ and $1$, respectively, and $\succeq$ stands for componentwise inequalities. Following the strategy proposed in \cite{Biouc2009,Iorda2012}, this problem can be solved using the alternating direction method of multipliers (ADMM)~\cite{Boyd2011}.
The optimization problem can be rewritten as
\begin{equation}\label{obj}
\begin{aligned}
	 &\min\limits_{\mathbf{U},\mathbf{V}}\frac{1}{2}\Vert\mathbf{V}_1-\mathbf{Y}\Vert^2_F +\lambda\Vert\mathbf{V}_3\Vert_{1,1}  +\iota_{\mathcal{N}}(\mathbf{V}_4) +\iota_{\mathcal{S}}(\mathbf{V}_5).\\
    & \text{s.t.}\
    \left\{
      \begin{array}{ll}
      \mathbf{V}_1&  =\mathbf{E}\mathbf{U}\\
      \mathbf{V}_2& =\mathbf{U}\\
      \mathbf{V}_3&=\mathbf{V}_2\mathbf{W}\\
    \end{array}
    \right.
    \ \text{and} \
        \left\{
      \begin{array}{ll}
      \mathbf{V}_4&=\mathbf{U}\\
       \mathbf{V}_5&=\mathbf{U}
    \end{array}
    \right.
\end{aligned}
\end{equation}
with $\mathbf{V} \triangleq \left[ \mathbf{V}_1\  \mathbf{V}_2\  \mathbf{V}_3\  \mathbf{V}_4\  \mathbf{V}_5 \right]$. By introducing
\begin{equation}\label{variable}
	\mathbf{G} = \begin{bmatrix} \mathbf{E} \\ \mathbf{I} \\ \mathbf{0} \\ \mathbf{I} \\ \mathbf{I} \end{bmatrix}^T
    \ \text{and}\ \
    \mathbf{B} = \begin{bmatrix} -\mathbf{I} & \mathbf{0} & \mathbf{0} & \mathbf{0} & \mathbf{0}
    \\  \mathbf{0} & -\mathbf{I} & \mathbf{W} & \mathbf{0} & \mathbf{0}
\\ \mathbf{0} & \mathbf{0} & -\mathbf{I} & \mathbf{0} & \mathbf{0}
\\ \mathbf{0} & \mathbf{0} & \mathbf{0} & -\mathbf{I} & \mathbf{0}
\\ \mathbf{0} & \mathbf{0} & \mathbf{0} & \mathbf{0} & -\mathbf{I} \end{bmatrix}.
\end{equation}
the problem can be rewritten as
\begin{equation}
	\min\limits_{\mathbf{U},\mathbf{V}} g(\mathbf{V}) \text{  subject to } \mathbf{GU}+\mathbf{VB}= \boldsymbol{0}.
\end{equation}
whose augmented Lagrangian is
\begin{equation}\label{obj2}
	 \mathcal{L}(\mathbf{U},\mathbf{V},\mathbf{D})=g(\mathbf{V})+\frac{\mu}{2}\Vert\mathbf{GU}+\mathbf{VB}-\mathbf{D}\Vert^2_F.
\end{equation}
where $\mu$ is a positive parameter and $\mathbf{D}=\left[\mathbf{D}_1,\ldots,\mathbf{D}_5\right]$ is the Lagrange multiplier. The general algorithmic scheme implemented to solve \eqref{obj2} is detailed in Algo. \ref{admm}, whose steps are detailed in the Appendix.\\

\begin{algorithm}[h!]
\caption{ADMM for the optimization problem \eqref{obj2}}\label{admm}
\begin{algorithmic}
\State \textbf{Input:} $\mathbf{Y}$, $\mathbf{E}$, $\mathbf{W}$
\State \textbf{Initialization:} $\mathbf{U}^{(0)},\mathbf{V}^{(0)},\mathbf{D}^{(0)}$
\While{not convergence}
\State $\mathbf{U}^{(k+1)} \leftarrow \text{arg}\min\limits_{\mathbf{U}} \mathcal{L}(\mathbf{U},\mathbf{V}^{(k)},\mathbf{D}^{(k)})$
\State $\mathbf{V}^{(k+1)}\leftarrow \text{arg}\min\limits_{\mathbf{V}} \mathcal{L}(\mathbf{U}^{(k+1)},\mathbf{V},\mathbf{D}^{(k)})$
\State $\mathbf{D}^{(k+1)}\leftarrow \text{arg}\min\limits_{\mathbf{D}} \mathcal{L}(\mathbf{U}^{(k+1)},\mathbf{V}^{(k+1)},\mathbf{D})$
\EndWhile
\State \textbf{Output:} $\mathbf{U}^{(k+1)}$, $\mathbf{V}^{(k+1)}$, $\mathbf{D}^{(k+1)}$
\end{algorithmic}
\end{algorithm}

When the regularizer $\phi(\cdots)$ is defined from the hyperspectral image \eqref{eq:weight_HI}, its first principal component \eqref{eq:weight_PC}, DSM \eqref{eq:weight_DSM} or their combinations \eqref{eq:weight_HI_DSM} and \eqref{eq:weight_PC_DSM}, the weighted difference operator $\mathbf{W}$ can be computed before the optimization process. However, the problem is more challenging when the weighted difference operator $\mathbf{W}$ is defined by a guidance map relying on the abundances, as in  \eqref{eq:weight_A} and \eqref{eq:weight_A_DSM}. In this work, one proposes to follow a strategy similar to the reweighted-$\ell_1$ minimization \cite{Cande2008}, which has been for instance used in ~\cite{He2017,zheng2016} to conduct sparse unmixing. It consists in alternatively updating the weighting operator $\mathbf{W}$ after each update of the abundance matrix $\mathbf{A}$. The optimization process is summarized in Algo. \ref{alg:algorithm3}.

\begin{algorithm}[h!]
\caption{Reweighted $\ell_1$-minimization}\label{alg:algorithm3}
\begin{algorithmic}
\State \textbf{Input:} $\mathbf{Y}$, $\mathbf{E}$, $\mathbf{W}$
\State \textbf{Initialization:} $\mathbf{W}$
\While{not convergence}
\State Updating $\mathbf{A}$ using Algo. \ref{admm}
\State Updating $\mathbf{W}$ with the new state $\mathbf{A}$
\EndWhile \label{euclidendwhile}
\State \textbf{Output:} $\mathbf{A}$
\end{algorithmic}
\end{algorithm}
% Data

\begin{figure*}
        \begin{subfigure}[b]{0.25\textwidth}
                \centering
                \includegraphics[width=.9\linewidth]{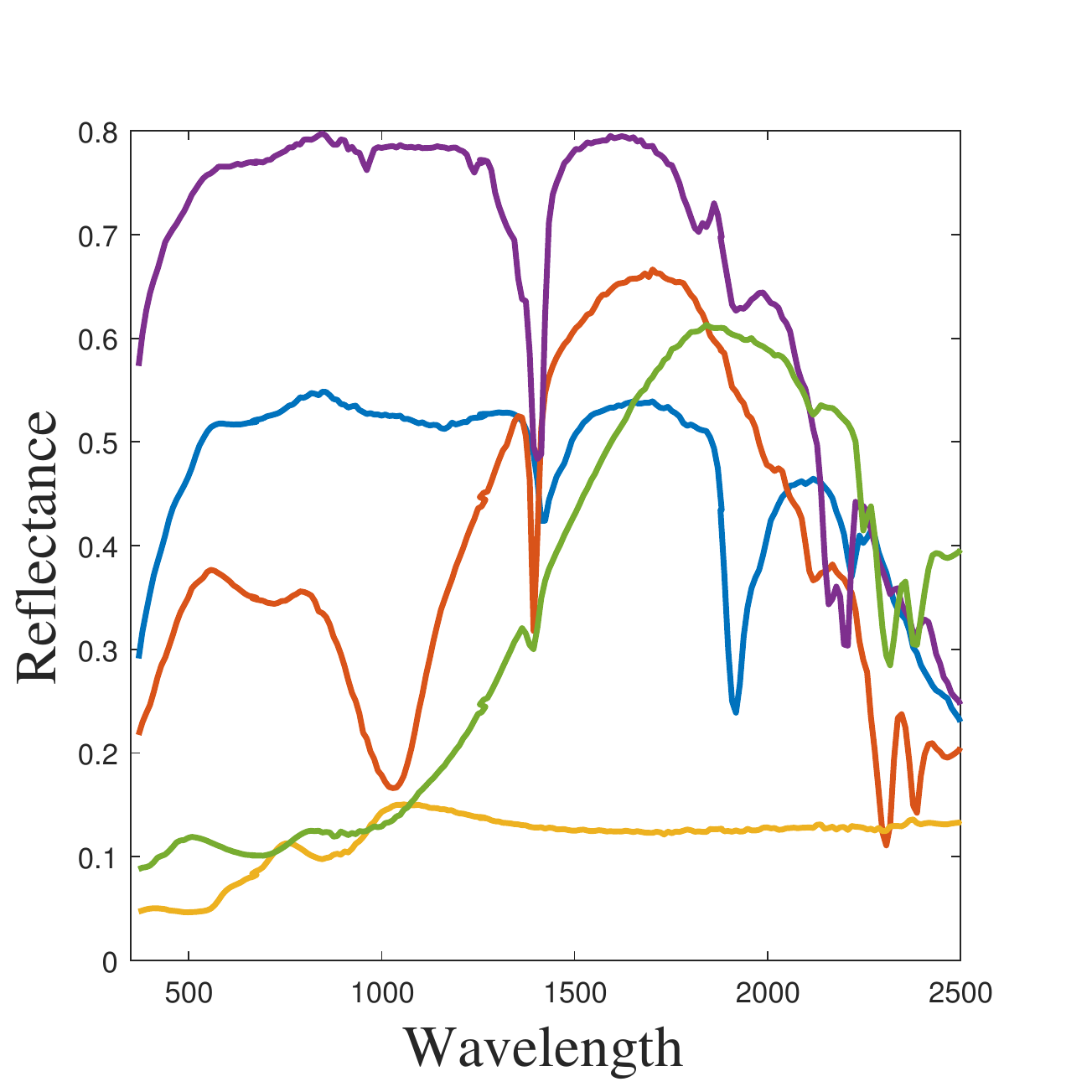}
                \caption{}
                \label{fig:endm_sim1}
        \end{subfigure}%
        \begin{subfigure}[b]{0.25\textwidth}
                \centering
                \includegraphics[width=.9\linewidth]{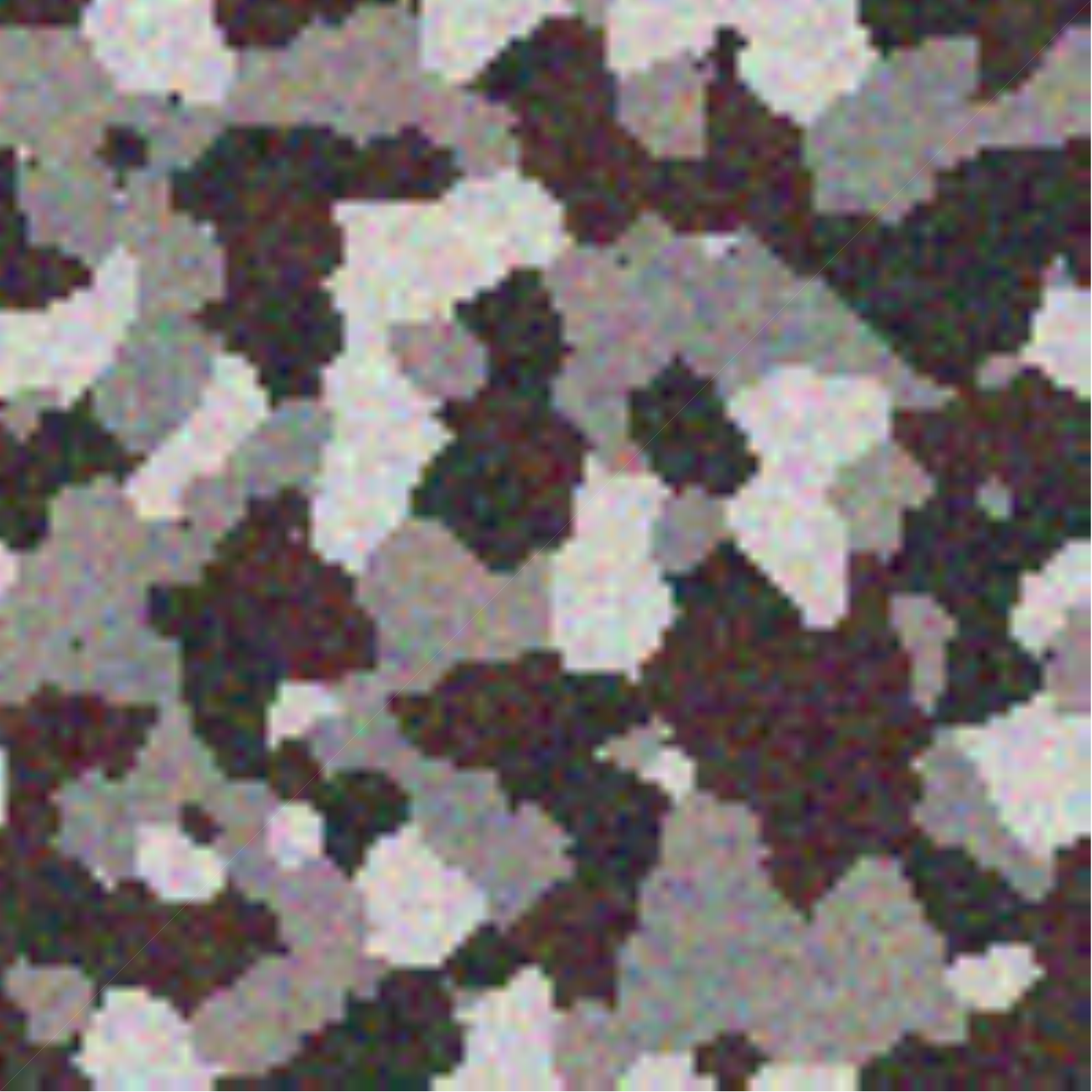}
                \caption{}
                \label{fig:hyp_sim1}
        \end{subfigure}%
        \begin{subfigure}[b]{0.25\textwidth}
                \centering
                \includegraphics[width=\linewidth]{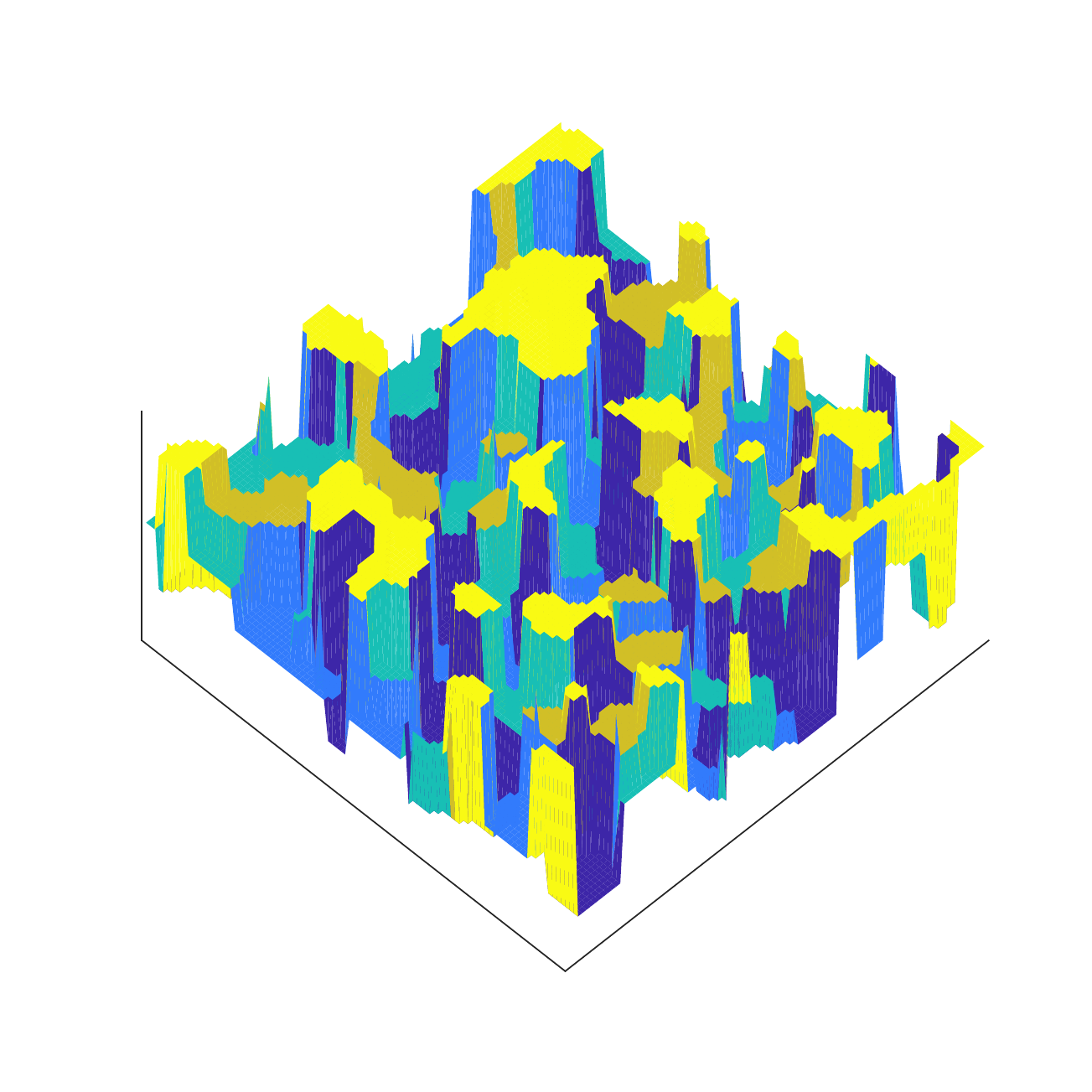}
                \caption{}
                \label{fig:lidar_sim1}
        \end{subfigure}%
        \begin{subfigure}[b]{0.25\textwidth}
                \centering
                \includegraphics[width=.9\linewidth]{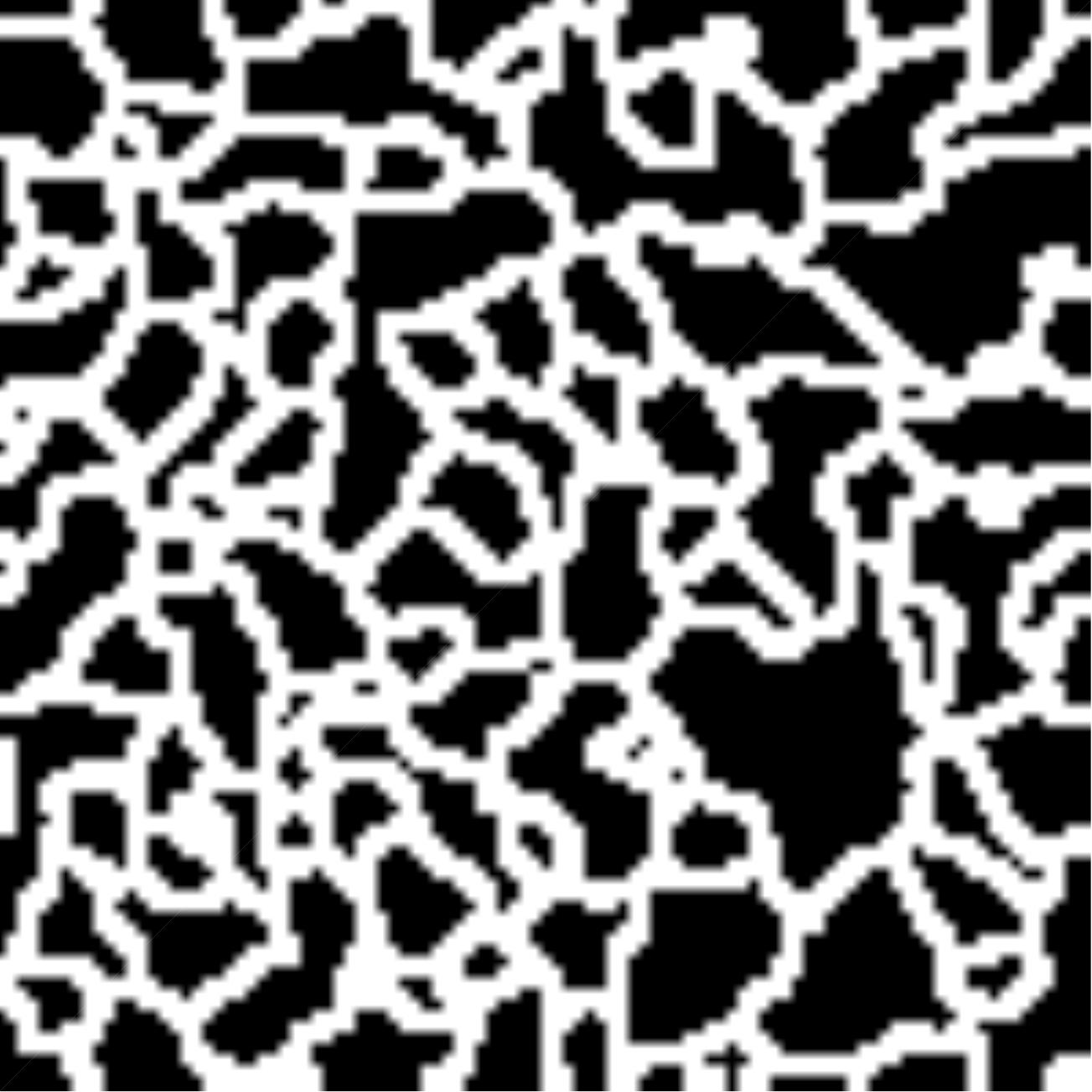}
                \caption{}
                \label{fig:grad_sim1}
        \end{subfigure}
        \caption{SIM1: (a) Endmember spectra used for simulated data. (b) Color composition of the synthetic hyperspectral image. (c) Synthetic DSM data. (d) Extracted edge pixels.}\label{fig:sim1}
\end{figure*}
% Results

\section{Experiments using simulated data}
To validate the proposed spectral unmixing algorithm, experiments are first conducted on simulated hyperspectral images. Two distinct synthetic datasets are considered, referred to as SIM1 and SIM2 in what follows. The first dataset (SIM1) relies on synthetically generated hyperspectral and DSM data while the second dataset (SIM2) combines a synthetically generated hyperspectral data with a real DSM. Their respective generation processes are described in the following paragraphs. For both datasets, quantitative validation has been conducted by computing the root mean square error (RMSE) associated with the estimated abundances, i.e.,
\begin{equation}
\mathrm{RMSE}_{\mathrm{w}} =\sqrt{\frac{1}{N M}\sum\limits_{i = 1}^{N}\sum\limits_{m = 1}^M{({{a}_{mi}} - {{\hat{a}}_{mi}} )}^2}.
\end{equation}
where ${a}_{mi}$ and $\hat{a}_{mi}$ are the actual and estimated abundance of the $m$th endmember in the $i$th pixel, respectively, and $N$ is the number of pixels in the whole image. The RMSE has been also computed for the pixels located on the edges between heterogeneous regions. In this case, RMSE is computed similarly as follows
\begin{equation}
\label{eq:RMSEe}
\mathrm{RMSE}_{\mathrm{e}} =\sqrt{\frac{1}{N_{\mathrm{e}} M}\sum\limits_{i = 1}^{N_{\mathrm{e}}}\sum\limits_{m = 1}^M{({{a}_{mi}} - {{\hat{a}}_{mi}} )}^2}.
\end{equation}
where $N_{\mathrm{e}}$ is the number of the pixels located on the edges. The identification of the edge areas in the two synthetic datasets will be conducted based on the available DSM.

\subsection{Synthetic hyperspectral and LiDAR data (SIM1)}
\subsubsection{Simulation protocol}
The first synthetic dataset is composed of a simulated hyperspectral image and simulated LiDAR measurements. To generate the hyperspectral image, $M=5$ endmember spectra have been randomly selected from the USGS spectral library (see Fig.~\ref{fig:endm_sim1}). Each endmember spectrum is composed of $L=224$ spectral bands ranging from visible and near-infrared (VNIR) to short-wave infrared (SWIR). The spatial mapping of these components is chosen according to a piecewise homogeneous distribution following a Potts-Markov model~\cite{Eches2011}. Within each homogeneous class, the abundances are randomly generated while ensuring the ANS and ASC constraints. The hyperspectral data is finally generated according to the LMM, corrupted by a white Gaussian noise corresponding to a signal-to-noise ratio (SNR) of $20$dB. A color composition of the resulting hyperspectral image is represented in Fig.~\ref{fig:hyp_sim1}. Moreover, the piecewise homogeneous distribution is also used to define a synthetic DSM, corrupted by a additive white Gaussian noise with SNR$=50$dB. This DSM is represented in Fig.~\ref{fig:lidar_sim1}. Note that, for brevity, results obtained from DSM with different SNRs ($40$dB and $30$dB) are not reproduced in the present manuscript but are reported in the supplementary document \cite{Uezato2017TR}. Finally, to compute RMSE in the edge areas following \eqref{eq:RMSEe}, the edges are automatically extracted by thresholding the gradient magnitude of the generated DSM, leading to the binary mask in Fig.~\ref{fig:grad_sim1}.

All compared methods require a parameter $\lambda$ controlling the spatial regularization. This parameter may greatly affect the accuracy of estimating abundances. To analyze the sensitivity of the methods with respect to (w.r.t.) this parameter, the quantitative figures-of-merit $\mathrm{RMSE}_{\mathrm{w}}$ and $\mathrm{RMSE}_{\mathrm{e}}$ are computed for $\lambda\in\left\{0.001, 0.05, 0.1, 0.5, 1, 1.5 \right\}$. The weighting functions introduced in Section \ref{section_weight} also require the parameters $\sigma_p^2$, $\sigma_y^2$, $\sigma_a^2$ and $\sigma_h^2$ controlling the weight range. They are chosen in the set $\sigma^2 \in \left\{10^{-5}, 10^{-4}, 0.001, 0.01, 0.1\right\}$. Unless otherwise stated, these parameters are   selected to the values which produce the lowest $\mathrm{RMSE}_{\mathrm{w}}$.

\begin{table}[h!]
\caption{SIM1: Abundance estimation errors for the whole and edge pixels using an optimal combination of $\sigma^2$ and $\lambda$.}\label{table:sim1_rmse}
%\resizebox{\columnwidth}{!}{
\centering
\begin{tabular}{cccccccc}			
\toprule			
& $\mathrm{RMSE}_{\mathrm{w}}$	& $\mathrm{RMSE}_{\mathrm{e}}$	\\
\midrule			
No-weight	&0.0165	&0.0165	\\
w-HI	&0.0088	&0.007	\\
w-PC1	&0.0097	&0.0077	\\
w-A	&0.0059	&0.0058	\\
w-DSM	&\textbf{0.0048}	&\textbf{0.0056}	\\
w-HI-DSM	&\textbf{0.0048}	&\textbf{0.0056}	\\
w-PC1-DSM	&0.0050	&0.0057	\\
w-A-DSM	&\textbf{0.0048}	&\textbf{0.0056}	\\
\bottomrule			
\end{tabular}												
%}
\end{table}

\subsubsection{Results and discussion}
Abundance errors $\mathrm{RMSE}_{\mathrm{w}}$ and  $\mathrm{RMSE}_{\mathrm{e}}$ computed for the whole pixels and the edge pixels are reported in  Table. \ref{table:sim1_rmse}.  The methods referred to as w-A, w-HI and w-PC1 which use the hyperspectral image, its first principal component or the abundances, respectively, to adjust the weight produce better abundance estimates than the no-weight method. As illustrated by the $\mathrm{RMSE}_{\mathrm{e}}$ reported in Table. \ref{table:sim1_rmse}, these three methods specifically lead to more accurate estimates of abundances for the pixels located in the edge. This demonstrates that the weights computed from these guidance maps  can correctly capture the spatial information and guide the regularization. Finally, the methods that incorporated the DSM information to design the weighted spatial regularization (i.e., w-DSM, w-A-DSM, w-HI-DSM and w-PC1-DSM) systematically improve abundance estimation w.r.t. their counterparts which do not benefit from the height information (see Table \ref{table:sim1_rmse}). Moreover, unlike the w-A, w-HI and w-PC1 methods, the methods incorporating DSM reach similar performance for the whole pixels and the pixels located in the edge areas.

\begin{figure}[h!]
\centering
        \begin{subfigure}[b]{0.45\textwidth}
                \centering
                \includegraphics[width=\linewidth]{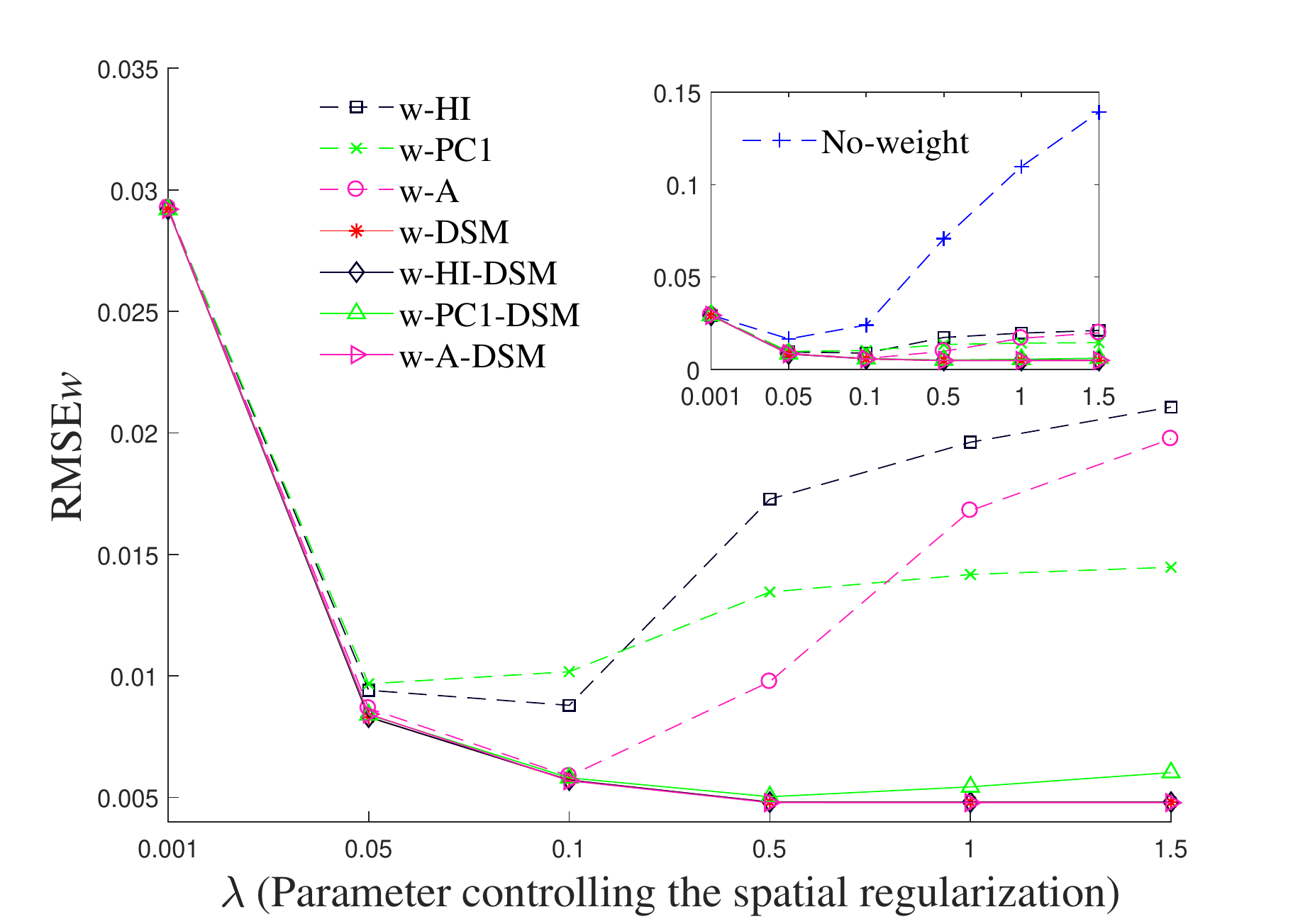}
                \caption{}
                \label{fig:rmse_a_sim1}
        \end{subfigure}%
        \\
        \begin{subfigure}[b]{0.45\textwidth}
                \centering
                \includegraphics[width=\linewidth]{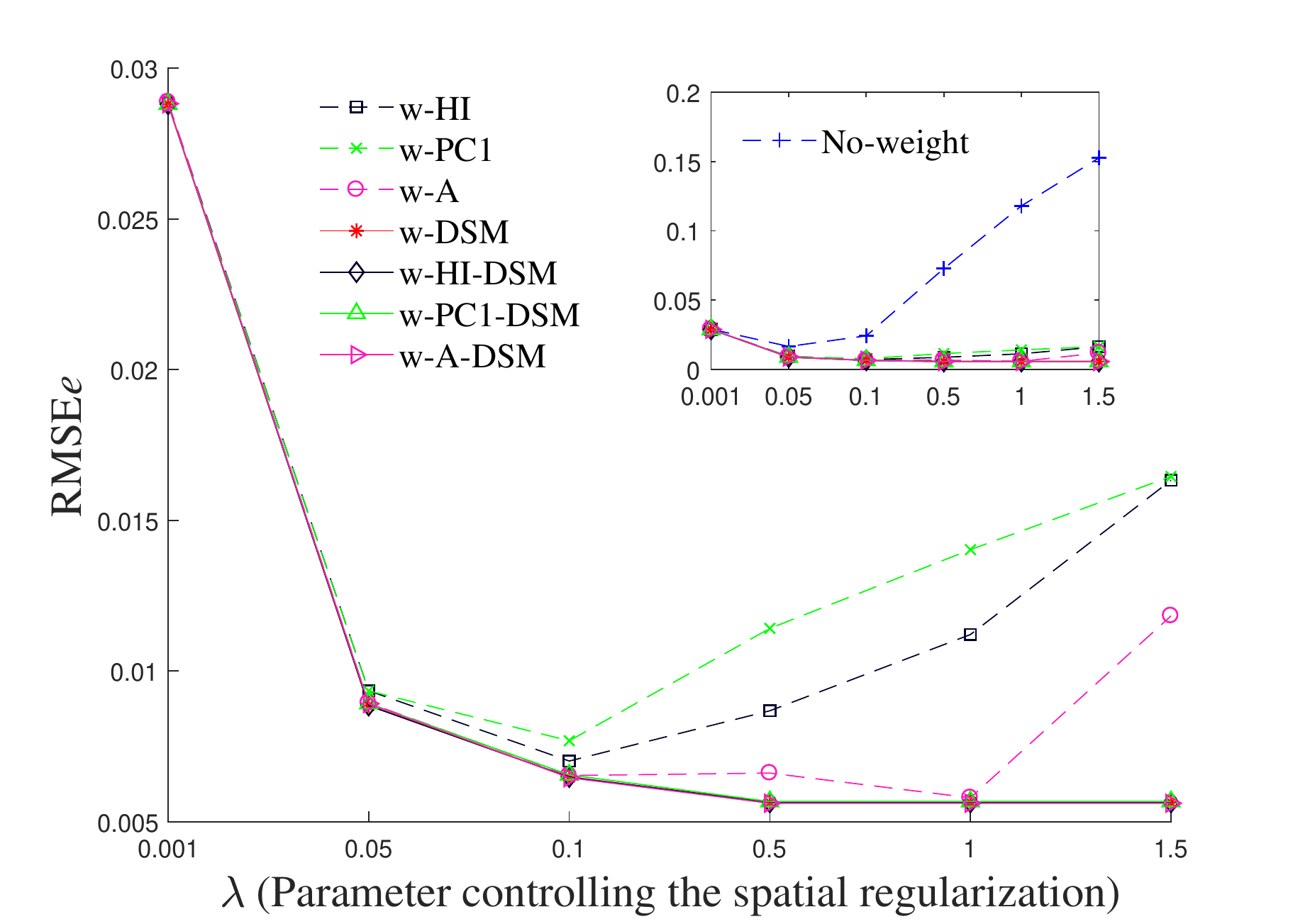}
                \caption{}
                \label{fig:rmse_a_edge_sim1}
        \end{subfigure}
        \caption{SIM1: Abundance estimation errors as functions of $\lambda$. (a) $\mathrm{RMSE}_{\mathrm{w}}$ computed for the whole pixels. (b) $\mathrm{RMSE}_{\mathrm{e}}$ computed for the pixels located in the  edge areas.}\label{fig:rmse_sim1}
\end{figure}

Besides, to evaluate the impact of the regularization parameter $\lambda$, the performances are represented as functions of $\lambda$ in Fig. \ref{fig:rmse_sim1}. The results clearly show that the method which does not include weighed regularization performs poorly compared with other methods. In particular, Fig. \ref{fig:rmse_sim1} shows that the absence of weighting leads to significant degradation performance for large values of $\lambda$. This implies that, in absence of weighted spatial regularization, it is much more challenging to choose an optimal parameter $\lambda$. Moreover, the methods exploiting the  DSM information are shown to be also more robust to varying values of the spatial regularization parameter than the DSM-free weighting methods. This shows that the spatial information provided by DSM can lead to robust estimates of abundances for a wide range of values of the regularization parameter $\lambda$.

% Data
\begin{figure*}
        \begin{subfigure}[b]{0.25\textwidth}
                \centering
                \includegraphics[width=.9\linewidth]{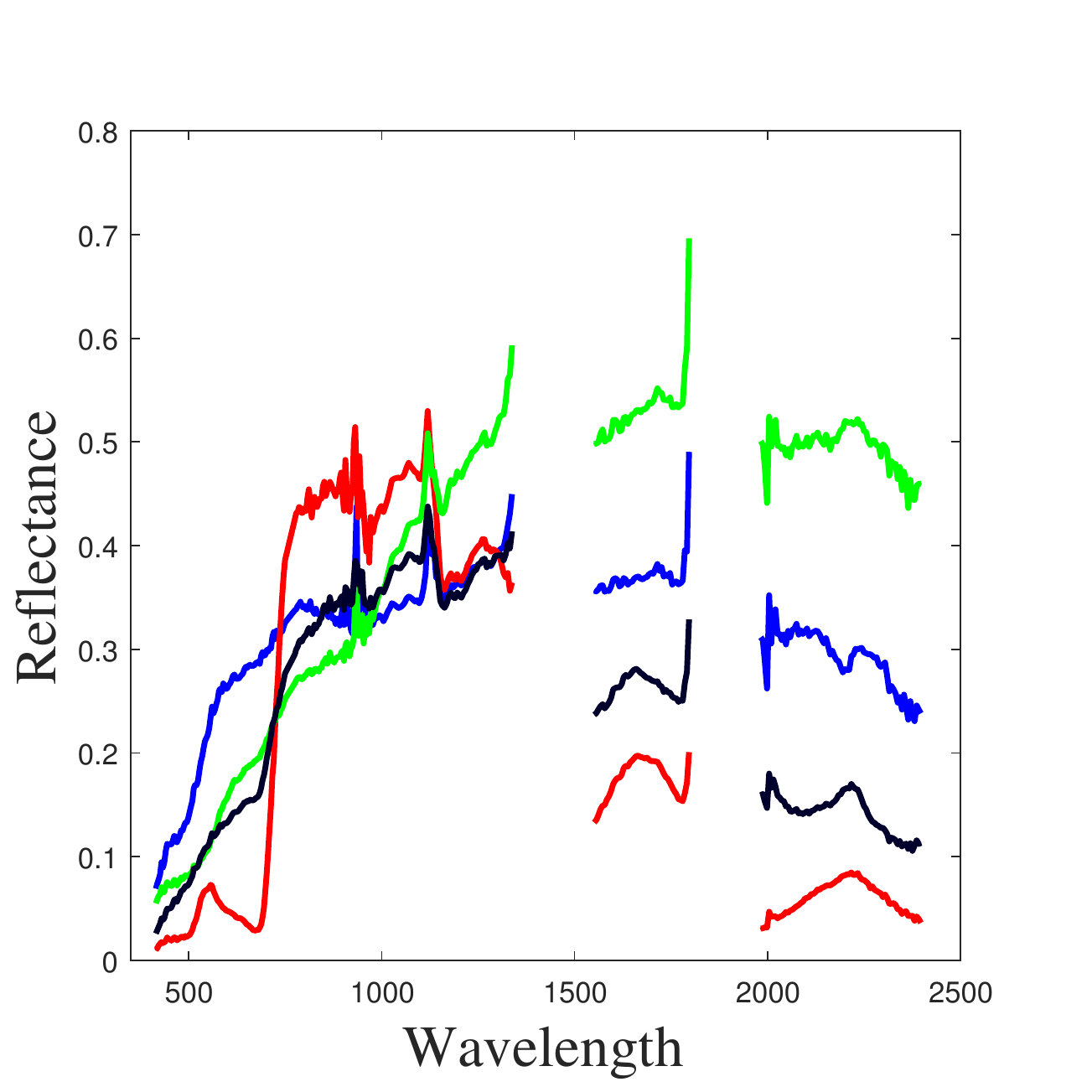}
                \caption{}
                \label{fig:endm_sim2}
        \end{subfigure}%
        \begin{subfigure}[b]{0.25\textwidth}
                \centering
                \includegraphics[width=.9\linewidth]{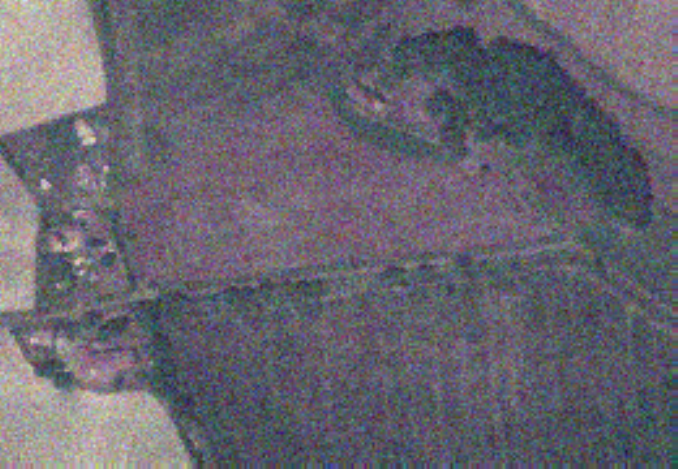}
                \caption{}
                \label{fig:hyp_sim2}
        \end{subfigure}%
        \begin{subfigure}[b]{0.25\textwidth}
                \centering
                \includegraphics[width=\linewidth]{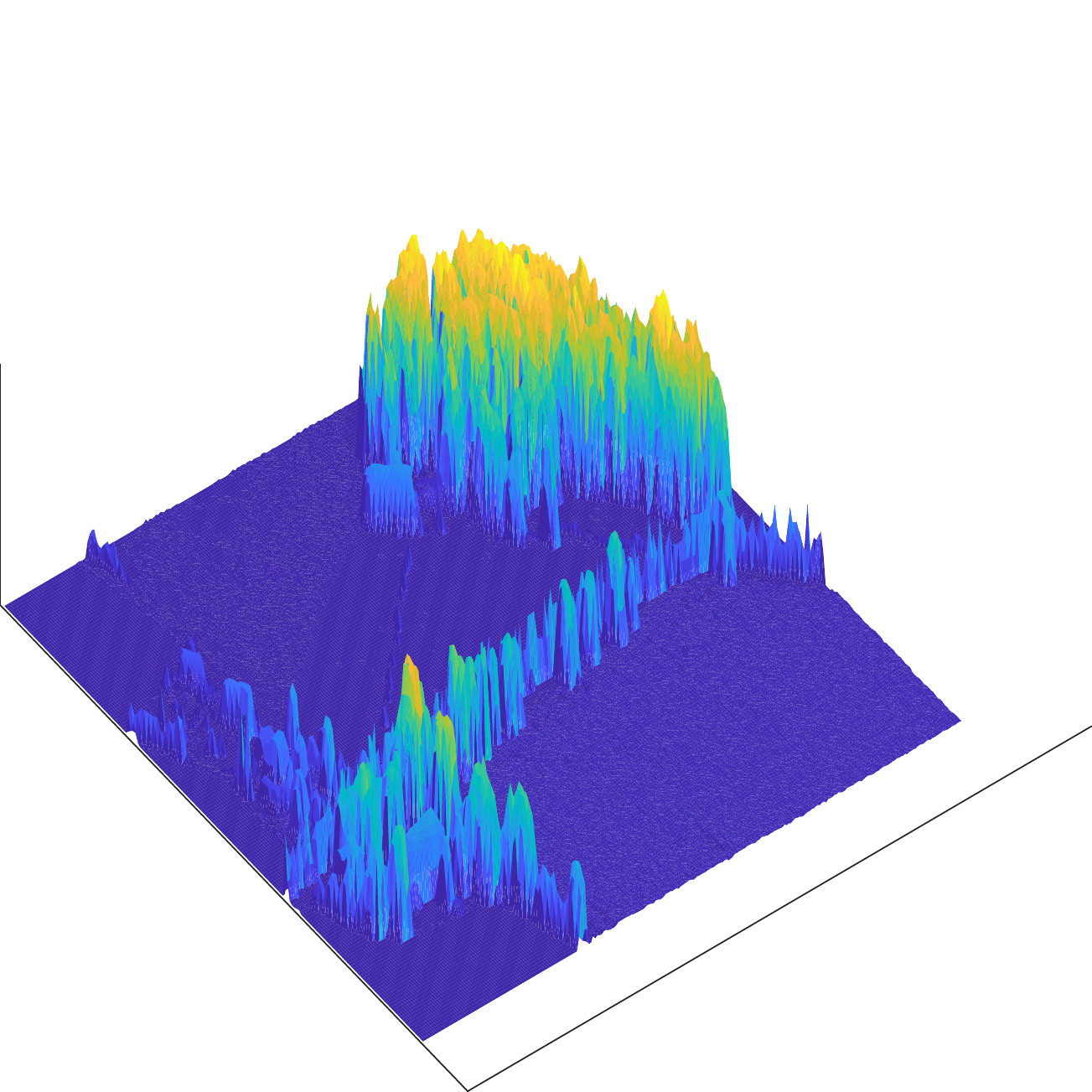}
                \caption{}
                \label{fig:lidar_sim2}
        \end{subfigure}%
        \begin{subfigure}[b]{0.25\textwidth}
                \centering
                \includegraphics[width=.9\linewidth]{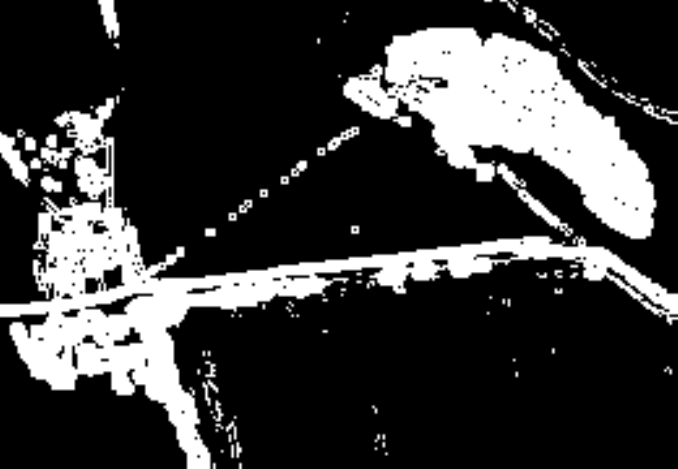}
                \caption{}
                \label{fig:grad_sim2}
        \end{subfigure}
        \caption{SIM2: (a) Endmember spectra used for simulated data. (b) Color composition image of the synthetic hyperspectral image. (c) Real DSM data. (d) Extracted edge pixels.}\label{fig:sim2}
\end{figure*}

\subsection{Synthetic hyperspectral data and real LiDAR data (SIM2)}
\label{subsec:SIM2}
\subsubsection{Simulation protocol}
In SIM1, simple spatially discrete piecewise homogeneous regions have been used to generate the simulated hyperspectral data and DSM. To provide complementary performance analysis on a more realistic scenario, a second synthetic dataset (referred to as SIM2) has been considered. First, four endmember spectra have been manually extracted from a real hyperspectral image composed of $260 \times 180$  pixels and acquired by the HySpex hyperspectral camera in June 2016 over Saint-Andr{\'e}, France. These spectral signatures are represented in Fig. \ref{fig:endm_sim2}, where spectral ranges $1.34-1.55\mu$m and $1.80-1.98\mu$m of poor SNR have been removed. A flat spectrum defined by a reflectance of $0.01$ for all bands is also considered as an additional shadow endmember. For this hyperspectral image, LiDAR data represented in Fig. \ref{fig:lidar_sim2} was simultaneously acquired. Thus, given these five endmembers,  the hyperspectral image has been unmixed with the w-DSM unmixing method, whose spatial regularization is weighted by DSM computed from the LiDAR data. Finally, a synthetic yet realistic hyperspectral image has been generated following the LMM with the five endmembers and the estimated abundance maps. An additive Gaussian noise corresponding to SNR$=20$dB is finally considered, leading to the hyperspectral image with the color composition shown in Fig. \ref{fig:hyp_sim2}. As for SIM1, a binary mask identifying the pixels located in edge areas is estimated by thresholding the gradient amplitude of DSM. This mask, shown in Fig.\ref{fig:grad_sim2}, will be used to specifically computed the $\mathrm{RMSE}_{\mathrm{e}}$ defined by \eqref{eq:RMSEe} for the edge pixels. Note that, when compared to SIM1, the SIM2 dataset is composed of real LiDAR measurements and more realistic abundance maps. Moreover, contrary to SIM1, DSM of SIM2 may not capture all edges between areas comprising the hyperspectral image.

As for SIM1, the performance of the unmixing procedures are evaluated in terms of RMSE computed for the pixels of the whole image or only for the pixels located in edge areas. For this dataset, the regularization parameter $\lambda$ is chosen in the set $\lambda \in \left\{0.001, 0.005, 0.01, 0.05, 0.1\right\}$ and the parameters $\sigma^2$ controlling the weight range have been selected in $\sigma^2 \in \left\{10^{-4}, 10^{-3}, 0.01, 0.1, 0.5, 1 \right\}$. Unless otherwise stated, these parameters are fixed to the values which produce the lowest $\mathrm{RMSE}_{\mathrm{w}}$.

\begin{table}[h!]
\centering
\caption{SIM2: Abundance estimation errors for the whole and edge pixels using an optimal combination of $\sigma^2$ and $\lambda$.}\label{table:sim2_rmse}
%\resizebox{\columnwidth}{!}{
\begin{tabular}{cccccccc}			
\toprule			
Method	& $\mathrm{RMSE}_{\mathrm{w}}$	& $\mathrm{RMSE}_{\mathrm{e}}$	\\
\midrule			
No-weight	&0.0146	&0.0189	\\
w-HI	&0.0143	&0.0185	\\
w-PC1	&0.0147	&0.0189	\\
w-A	&0.0136	&0.017	\\
w-DSM	&\textbf{0.0125}	&0.0154	\\
w-HI-DSM	&\textbf{0.0125}	&0.0154	\\
w-PC1-DSM	&\textbf{0.0125}	&0.0155	\\
w-A-DSM	&\textbf{0.0125}	&\textbf{0.0151}\\
\bottomrule			
\end{tabular}															
%}
\end{table}

\subsubsection{Results and discussion}
As shown in Table \ref{table:sim2_rmse}, the four methods w-DSM, w-PC1-DSM, w-HI-DSM and w-A-DSM exploiting the availability of DSM provides relatively smaller abundance estimation error $\mathrm{RMSE}_{\mathrm{w}}$ and $\mathrm{RMSE}_{\mathrm{e}}$ than the other methods.

\begin{figure}[h!]
\centering
        \begin{subfigure}[b]{0.45\textwidth}
                \centering
                \includegraphics[width=.9\linewidth]{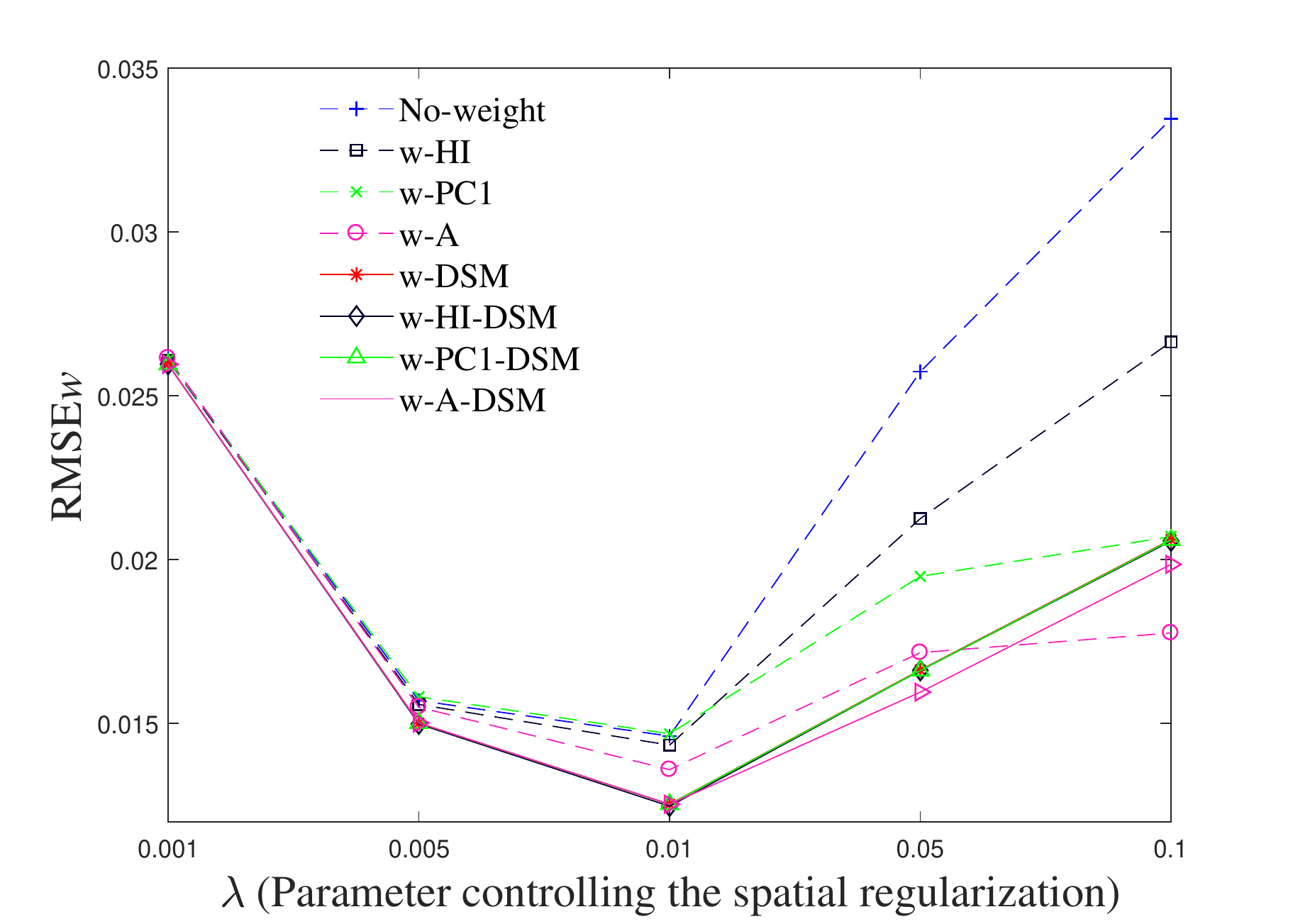}
                \caption{}
                \label{fig:rmse_a_sim2}
        \end{subfigure}%
        \\
        \begin{subfigure}[b]{0.45\textwidth}
                \centering
                \includegraphics[width=.9\linewidth]{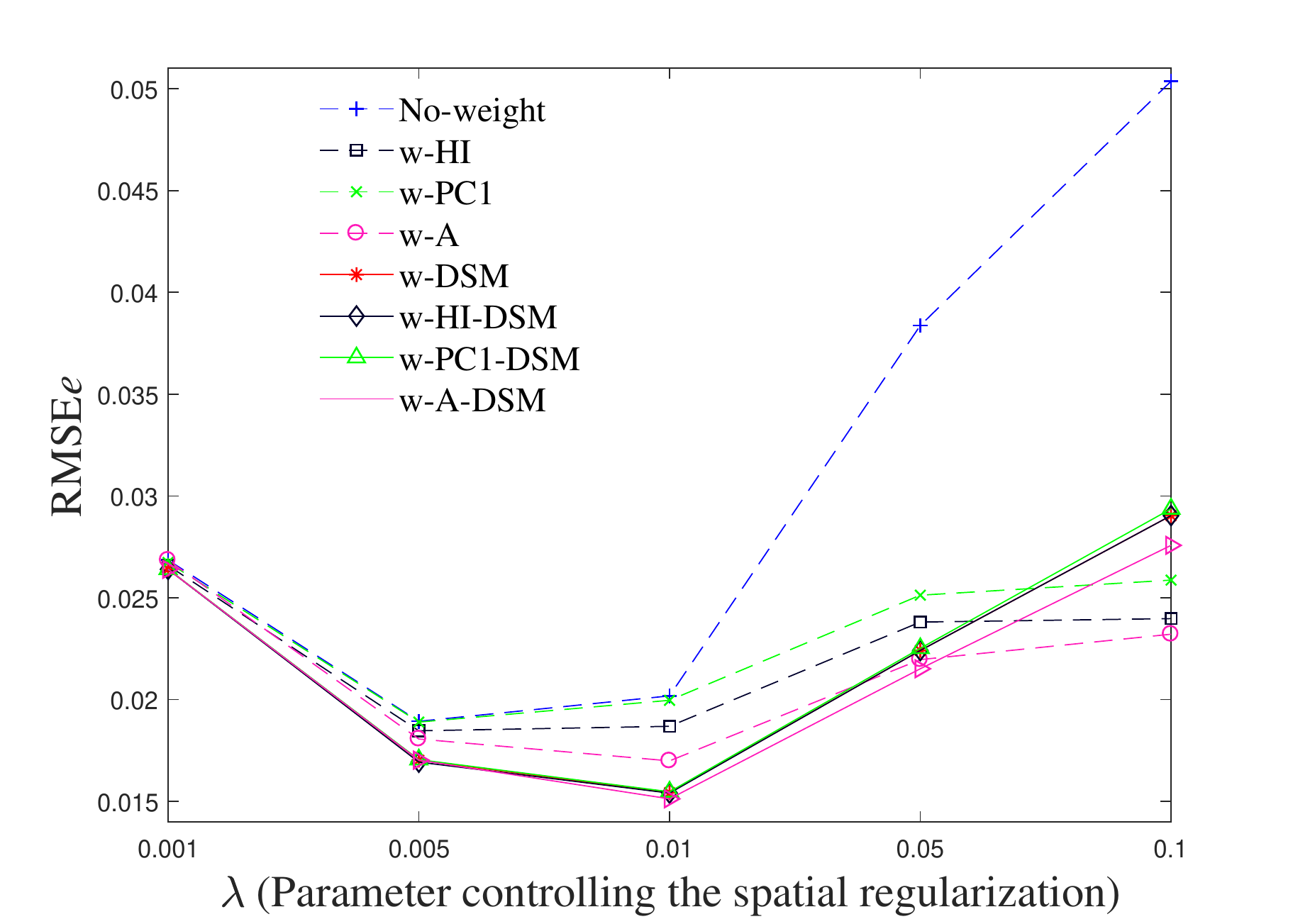}
                \caption{}
                \label{fig:rmse_a_edge_sim2}
        \end{subfigure}
        \caption{SIM2: Abundance estimation errors as functions of $\lambda$. (a) $\mathrm{RMSE}_{\mathrm{w}}$ computed for the whole pixels. (b) $\mathrm{RMSE}_{\mathrm{e}}$ computed for the pixels located in the  edge areas.}\label{fig:rmse_sim2}
\end{figure}

Similarly to the behavior encountered for SIM1, the performance of the method which does not use a weighted spatial regularization significantly degrades when the parameter $\lambda$ increases, as shown in Fig. \ref{fig:rmse_sim2}. This behavior is even worse when focusing specifically on the edge areas. On the other hand, the unmixing methods incorporating the height information provided by DSM lead to smaller $\mathrm{RMSE}_{\mathrm{w}}$ and $\mathrm{RMSE}_{\mathrm{e}}$ for a wide range of the regularization parameter, except for $\lambda =0.1$. This can be explained by the fact that, for this dataset composed of real LiDAR measurements, changes in the ground composition (i.e., abundance maps) do not systematically lead to changes in DSM. Thus, for high value of $\lambda$, the DSM-based unmixing method tends to promote spatial coherence of the abundance maps  in edge areas erroneously, as illustrated in Fig. \ref{fig:rmse_sim2}(b).

% Result

\begin{figure*}
        \begin{subfigure}[b]{0.25\textwidth}
                \centering
                \includegraphics[width=.9\linewidth]{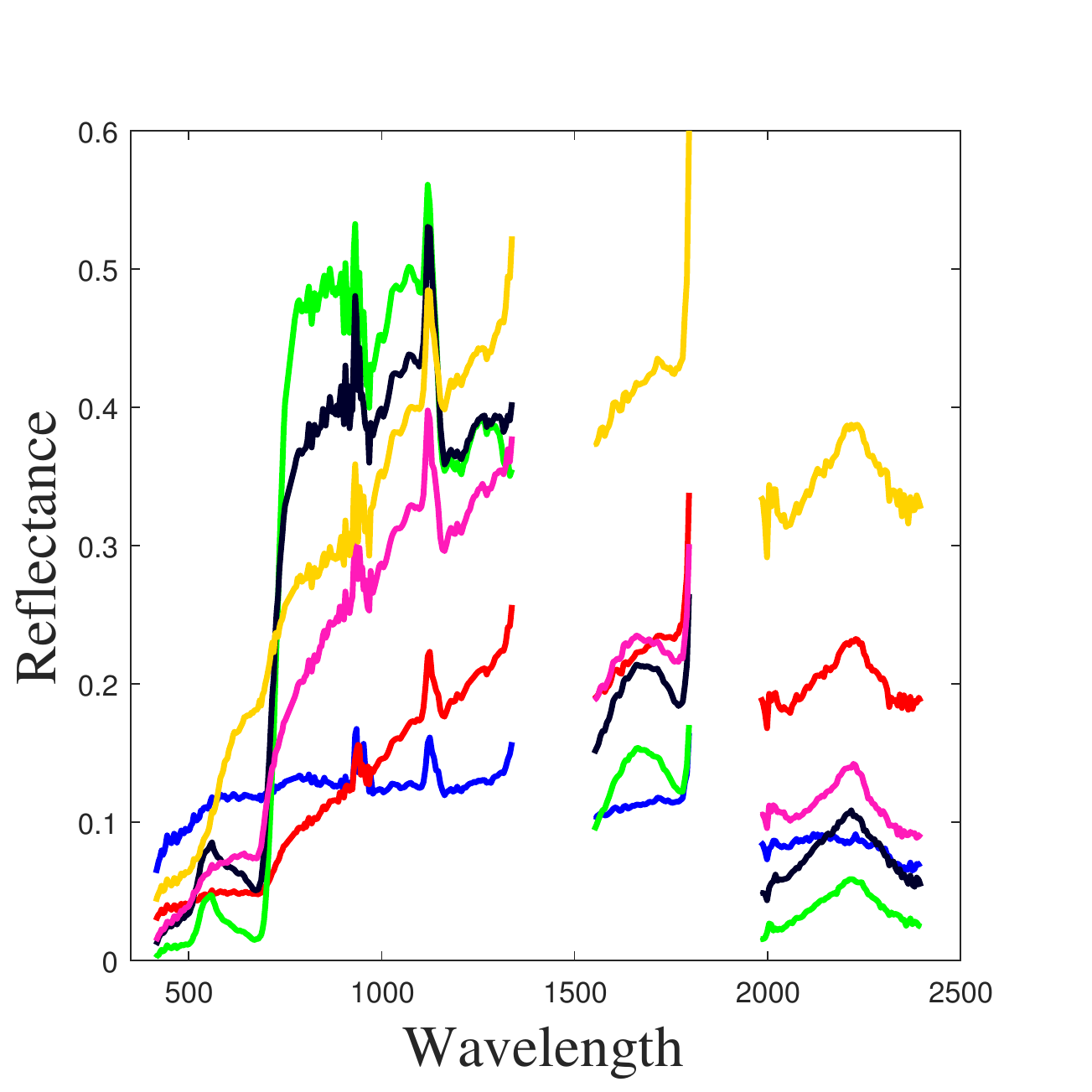}
                \caption{}
                \label{fig:endm_real}
        \end{subfigure}%
        \begin{subfigure}[b]{0.25\textwidth}
                \centering
                \includegraphics[width=.9\linewidth]{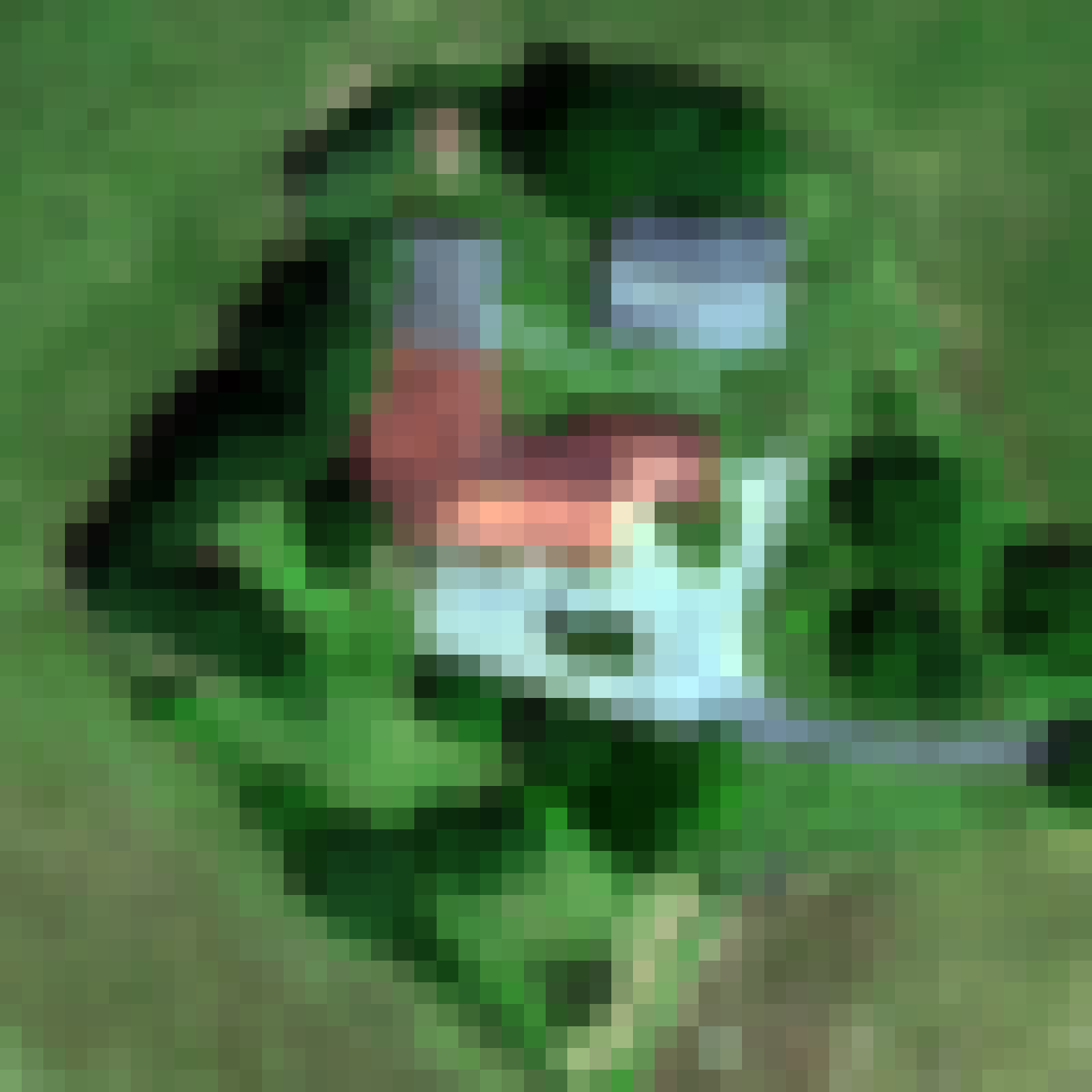}
                \caption{}
                \label{fig:hyp_real}
        \end{subfigure}%
        \begin{subfigure}[b]{0.25\textwidth}
                \centering
                \includegraphics[width=\linewidth]{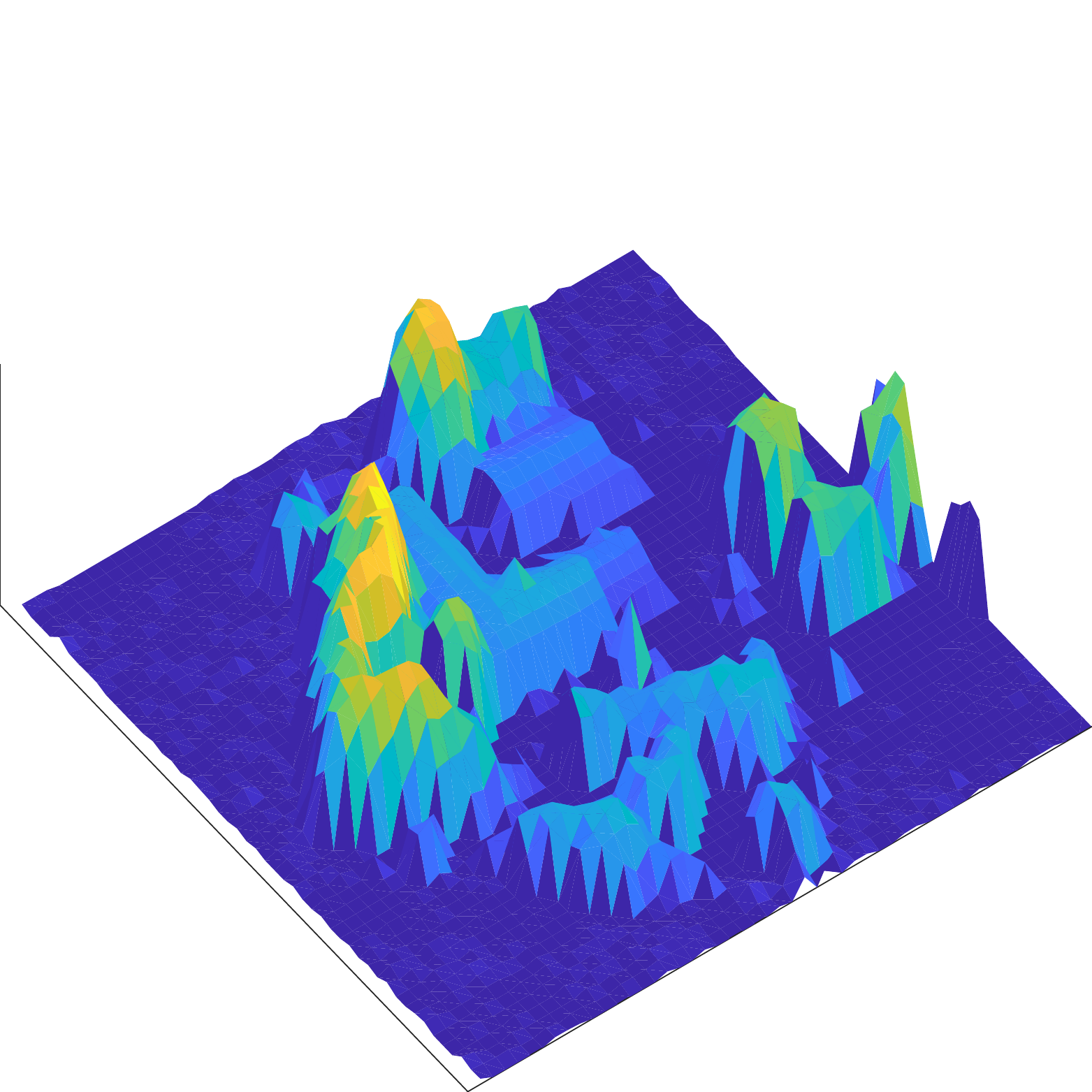}
                \caption{}
                \label{fig:lidar_real}
        \end{subfigure}%
        \begin{subfigure}[b]{0.25\textwidth}
                \centering
                \includegraphics[width=.9\linewidth]{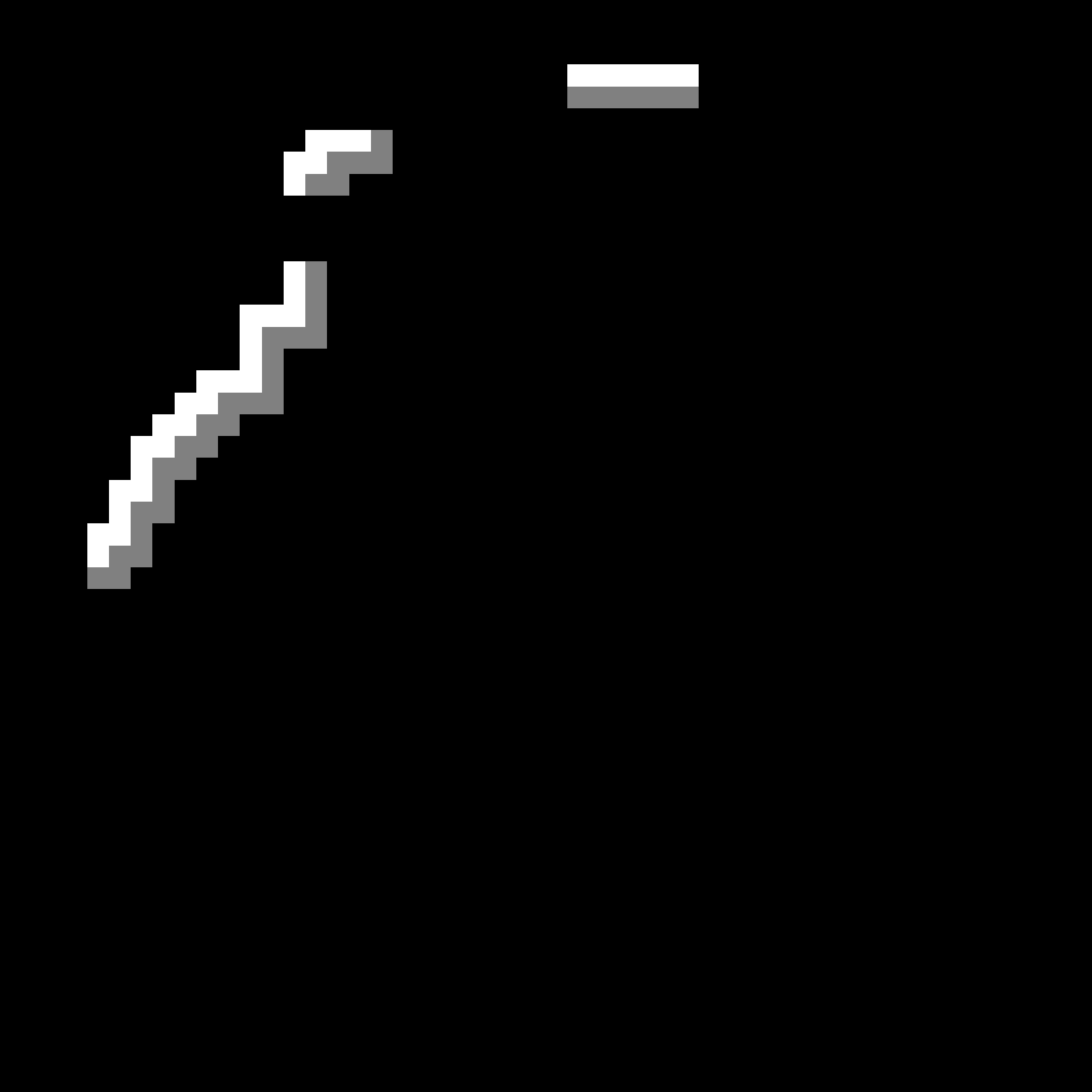}
                \caption{}
                \label{fig:tree_pixels}
        \end{subfigure}
        \caption{Real image: (a) Endmember spectra extracted from the hyperspectral image. (b) Color composition of the real hyperspectral image. (c) Real DSM data. (d) Pixels located on the boundary of trees (white) and shaded grass (gray).}\label{fig:real_im}
\end{figure*}

\section{Experiments using real hyperspectral and LiDAR data}
\subsection{Description of the dataset and experimental protocol}
The real hyperspectral image and real LiDAR data, already used to generate the synthetic dataset SIM2 in Section \ref{subsec:SIM2}, were acquired in June 2016, over the city of Saint-Andr{\'e}, France. The hyperspectral image was composed of $415$ spectral bands ranging from VNIR to SWIR ($0.40-2.40\mu$m). The spectral bands in the spectral ranges $1.34-1.55\mu$m and $1.80-1.98\mu$m have been removed since the bands were strongly affected by a large amount of noise. A $50 \times 50$ subset of the hyperspectral image has been extracted from the whole image and is depicted in Fig. \ref{fig:hyp_real}. Note that this scene of interest includes pixels of shadow, represented by dark pixels in the color composition, due to the presence of trees.

Endmember spectra of six distinct materials (i.e., tree, grass, soil, road, building 1 and building 2) have been  manually extracted based on prior knowledge of the scene. An additional endmember corresponding to shadow has been also considered to account for possible illumination variations and mitigate the presence of shadow. Unlike the simulated datasets SIM1 and SIM2, ground truth in terms of actual abundance maps is not available for this real hyperspectral image. Thus, the unmixing performances of the algorithms are qualitatively evaluated thanks to visual inspection. More precisely, this experiment has been designed to assess whether DSM can help reducing the impact of shadow pixels on the abundance estimation. Indeed, for the hyperspectral image corrupted by shadows, guidance maps derived from the hyperspectral image, its principal components or abundances are expected to be affected by these illumination variations. This may lead to erroneous spatial information incorporated into the corresponding weighted spatial regularizations. Conversely, height information provided by external LiDAR data under the form of DSM is known to be insensitive to presence of shadows in the scene, which may produce more accurate guidance maps. To locate pixels possibly affected by these effects, the areas corresponding to trees and the shaded grass are manually identified  by visually inspecting DSM and the color composition. These pixels, shown in Fig. \ref{fig:tree_pixels}, will be of particular interest to evaluate the consistence of the estimated abundance maps.

\subsection{Results and discussion}
The abundance maps associated with the distinct materials have been estimated by the unmixing spatially regularized unmixing methods with the parameters $\sigma^2$ and $\lambda$ empirically tuned to provide the less sensitive abundance maps w.r.t. the shadow effects. The estimated abundance maps of the buildings, soil and road are similar for all the methods and are not reproduced here for brevity (they can be found in the associated technical report \cite{Uezato2017TR}).

\begin{figure}[h!]
    \centering
    \includegraphics[width=\linewidth]{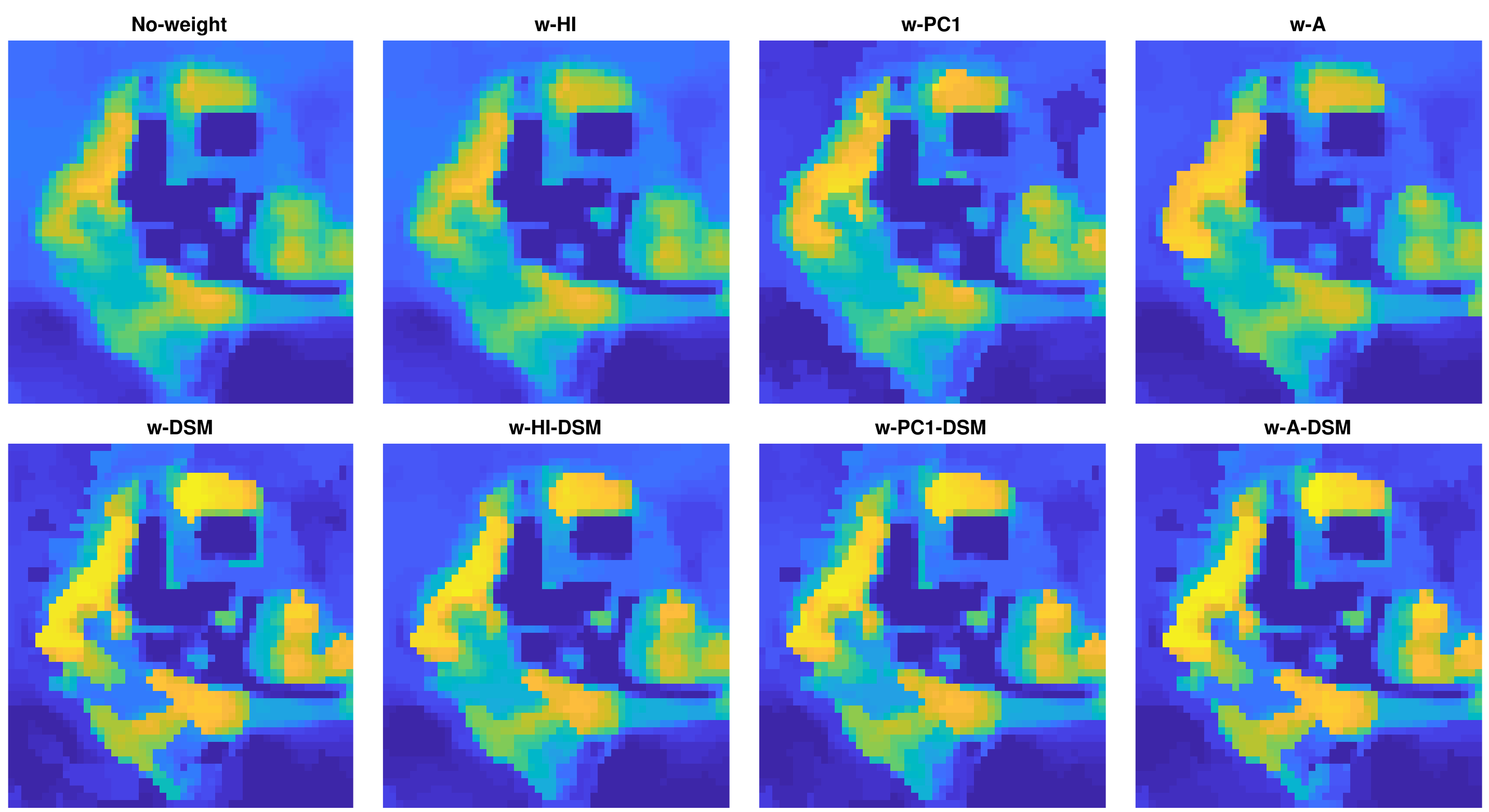}
    \caption{Real image: abundances estimated for tree.}
    \label{fig:abun_2class_tree}
\end{figure}

\begin{figure}[h!]
    \centering
    \includegraphics[width=\linewidth]{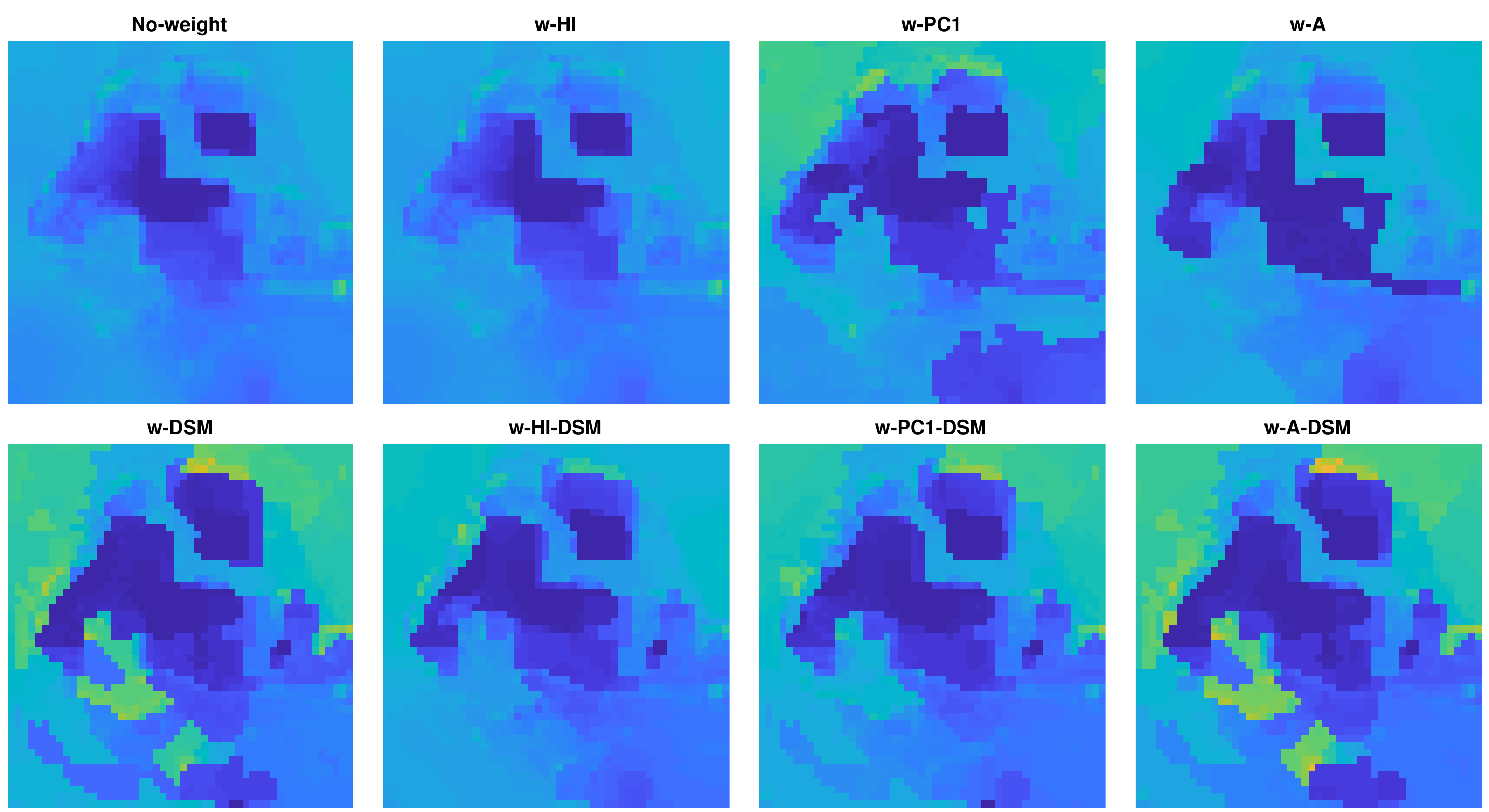}
    \caption{Real image: abundances estimated for grass.}
    \label{fig:abun_2class_grass}
\end{figure}

\begin{figure}[h!]
    \centering
    \includegraphics[width=\linewidth]{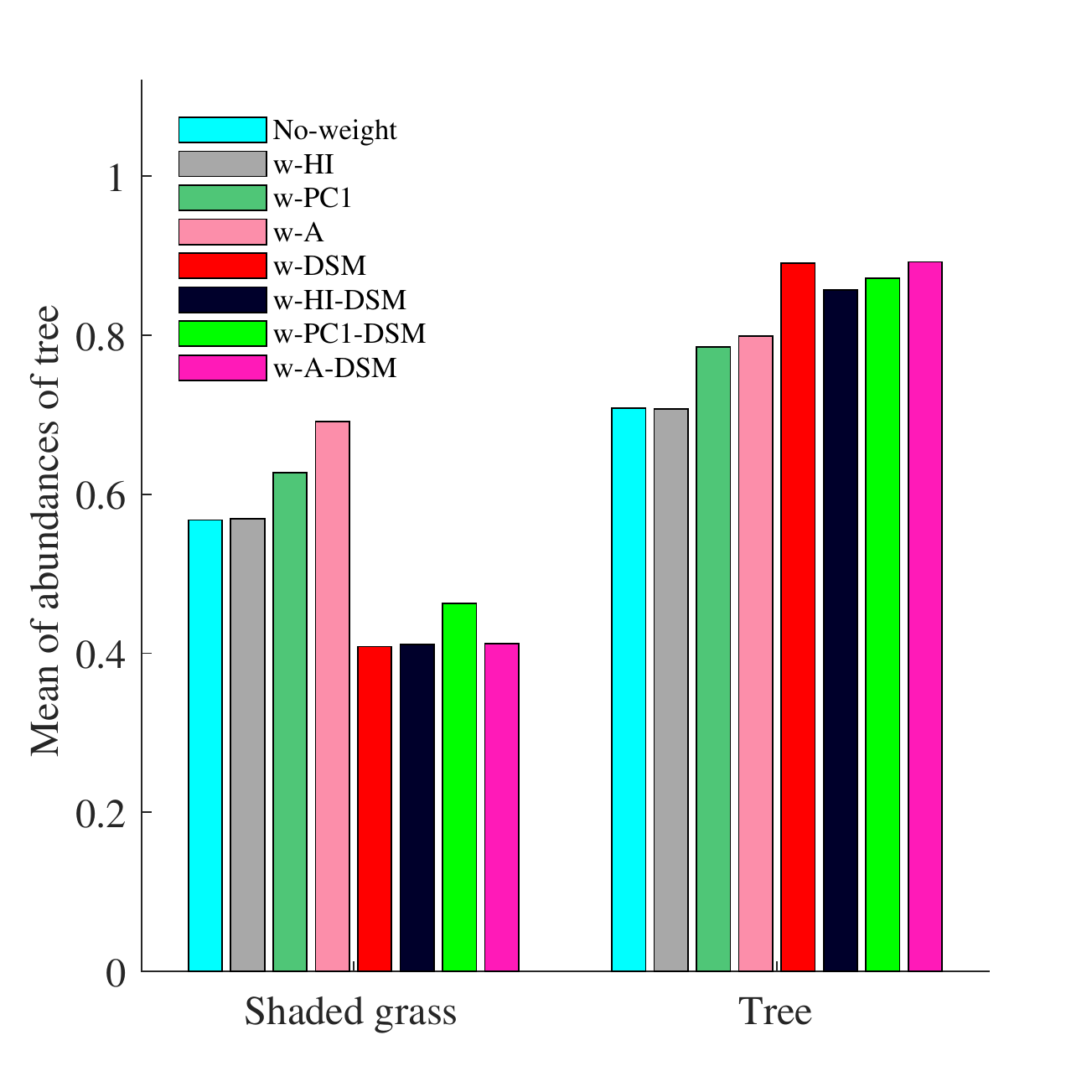}
    \caption{Average grass (left) and tree (right) abundances of tree pixels located on the boundary between grass and tree.}
    \label{fig:abun_tree}
\end{figure}

However, noticeable differences are observed for the abundance maps of the tree and the grass estimated by the methods incorporating or ignoring DSM. Indeed, as illustrated in Fig. \ref{fig:abun_2class_tree} and  Fig. \ref{fig:abun_2class_grass}, the abundances estimated by the no-weight, w-A, w-HI and w-PC1 methods are significantly affected by the shadow and show smooth transitions between pixels fully composed of tree and those fully composed of grass. These smooth transitions may be explained by nonlinear interactions due to scattering effects between the trees and the grass, often observed in vegetated areas \cite{Dobig2014,Heyle2014}. Thus, when computing the regularization weights directly from the hyperspectral image, the spatial information may be corrupted and not correctly adjust the spatial regularization. To quantitatively illustrate this finding, the means of estimated tree abundances of the pixels located in the boundary between shaded grass and tree have been computed for each method. The results are reported in Fig. \ref{fig:abun_tree}. The four methods ignoring DSM information (i.e., no-weight, w-A, w-HI and w-PC1) seem to overestimate the tree abundances in the pixels of grass affected by shadow while underestimating these abundances in the pixels of tree. Conversely, the abundances estimated by the methods incorporating DSM-based weighted regularization are shown to be less sensitive to the shadow since leading to more sharper abundance maps.

\section{Conclusion}
This paper proposed a general framework to incorporate external DSM information into spatially regularized spectral unmixing algorithms. Spatial information was derived from (possibly combined) guidance maps and was exploited to weight the spatial regularization accordingly. The performances of methods using  a unique guidance image or a combination of them were compared using simulated and real datasets, composed of a hyperspectral image and LiDAR data. These experiments showed that weighting the spatial regularization using a guidance map consistently outperformed weight-free spatial regularization. Moreover, when available, DSM allowed a complementary guidance map to be easily designed, which led to more accurate abundance estimates than the ones obtained by the DSM-free counterpart methods. This paper deeply focused on TV-like  regularizations but the proposed methodology has a great potential to be instanced with other spatial regularizations. These opportunities will be investigated in future works.

\appendix
%\section{}
This appendix provides details on the resolution of the optimization problem tackled by Algorithm \ref{admm}. The proposed approach is similar to \cite{Iorda2012}, with the difference that the proposed method imposes the sum-to-one constraint instead of imposing a sparsity constraint within each pixel.

First, the optimization w.r.t. $\mathbf{U}$ is written as
\begin{equation*}
	\begin{aligned}
	\mathbf{U}^{(k+1)} & = \textnormal{arg}\min\limits_{\mathbf{U}}\frac{\mu}{2}\Vert\mathbf{E}\mathbf{U}-\mathbf{V}_1^{(k)}-\mathbf{D}_1^{(k)}\Vert^2_F\\
    & +\frac{\mu}{2}\Vert\mathbf{U}-\mathbf{V}_2^{(k)}-\mathbf{D}_2^{(k)}\Vert^2_F+\frac{\mu}{2}\Vert\mathbf{U}-\mathbf{V}_4^{(k)}-\mathbf{D}_4^{(k)}\Vert^2_F\\
    & +\frac{\mu}{2}\Vert\mathbf{U}-\mathbf{V}_5^{(k)}-\mathbf{D}_5^{(k)}\Vert^2_F.\\
    \end{aligned}
\end{equation*}
which is solved by
\begin{equation*}
	\begin{aligned}
	\mathbf{U}^{(k+1)} & = \left( \mathbf{E}^T\mathbf{E}+3\mathbf{I}\right)^{-1}\left(\mathbf{E}^T\mathbf{F}^{(k)}_1+\mathbf{F}^{(k)}_2+\mathbf{F}^{(k)}_4+\mathbf{F}^{(k)}_5 \right).\\
    \end{aligned}
\end{equation*}
with
\begin{equation*}
\left\{
\begin{array}{ll}
  \mathbf{F}^{(k)}_1&=\mathbf{V}_1^{(k)}+\mathbf{D}_1^{(k)}\\
  \mathbf{F}^{(k)}_2&=\mathbf{V}_2^{(k)}+\mathbf{D}_2^{(k)}
\end{array}\right.
\ \text{and} \
\left\{
\begin{array}{rl}
  \mathbf{F}^{(k)}_4&=\mathbf{V}_4^{(k)}+\mathbf{D}_4^{(k)}\\
  \mathbf{F}^{(k)}_5&=\mathbf{V}_5^{(k)}+\mathbf{D}_5^{(k)}.
\end{array}\right.
\end{equation*}

The optimization w.r.t. $\mathbf{V}_1$ consists in solving
\begin{equation*}
	\mathbf{V}^{(k+1)}_1 = \textnormal{arg}\min\limits_{\mathbf{V}_1}\frac{1}{2}\Vert\mathbf{V}_1-\mathbf{Y}\Vert^2_F+\frac{\mu}{2}\Vert\mathbf{E}\mathbf{U}^{(k)}-\mathbf{V}_1-\mathbf{D}_1^{(k)}\Vert^2_F.
\end{equation*}
which can be conducted by
\begin{equation*}
	\mathbf{V}^{(k+1)}_1 = \frac{1}{1+\mu}\left[\mathbf{Y}+\mu\left(\mathbf{E}\mathbf{U}^{(k)}-\mathbf{D}_1^{(k)}\right)\right].
\end{equation*}

The optimization problem associated with $\mathbf{V}_2$ is
\begin{equation*}
	\begin{aligned}
	\mathbf{V}^{(k+1)}_2 = \textnormal{arg}\min\limits_{\mathbf{V}_2} & \frac{\mu}{2}\Vert\mathbf{U}^{(k)}-\mathbf{V}_2-\mathbf{D}_2^{(k)}\Vert^2_F\\
    & +\frac{\mu}{2}\Vert\mathbf{V}_2\mathbf{W}-\mathbf{V}_3^{(k)}-\mathbf{D}_3^{(k)}\Vert^2_F.
    \end{aligned}
\end{equation*}
and the resulting updating rule is
\begin{equation*}
	\mathbf{V}^{(k+1)}_2 = \left(\mathbf{U}^{(k)}-\mathbf{D}_2^{(k)}+\varphi_3\mathbf{W}^T\right)\left(\mathbf{I}+\mathbf{W}\mathbf{W}^T\right)^{-1}.
\end{equation*}
where $\mathbf{F}^{(k)}_3=\mathbf{V}_3^{(k)}+\mathbf{D}_3^{(k)}$.

Optimizing w.r.t. $\mathbf{V}_3$ consists in solving
\begin{equation*}
	\mathbf{V}^{(k+1)}_3 = \textnormal{arg}\min\limits_{\mathbf{V}_3} \frac{\mu}{2}\Vert\mathbf{V}_2^{(k)}\mathbf{W}-\mathbf{V}_3-\mathbf{D}_3^{(k)}\Vert^2_F + \Vert\mathbf{V}_3\Vert_{1,1}.
\end{equation*}
which is a standard sparse regression problem solved by a soft-thresholding step \cite{Combe2005}
\begin{equation*}
	\mathbf{V}^{(k+1)}_3 = \textnormal{soft}\left(\mathbf{V}_2^{(k)}\mathbf{W}-\mathbf{D}_3^{(k)},\frac{\lambda}{\mu}\right).
\end{equation*}

The optimization w.r.t. $\mathbf{V}_4$ is written
\begin{equation*}
	\mathbf{V}^{(k+1)}_4 = \textnormal{arg}\min\limits_{\mathbf{V}_4} \frac{\mu}{2}\Vert\mathbf{U}^{(k)}-\mathbf{V}_4-\mathbf{D}_4^{(k)}\Vert^2_F + \iota_{\mathcal{N}}(\mathbf{V}_4).
\end{equation*}
and solved by the projection
\begin{equation*}
	\mathbf{V}^{k+1}_4 = \textnormal{max}\left(\mathbf{U}^{(k)}-\mathbf{D}_4^{(k)},0\right).
\end{equation*}

Optimizing w.r.t. $\mathbf{V}_5$ consists in solving
\begin{equation*}
	\mathbf{V}^{(k+1)}_5 = \textnormal{arg}\min\limits_{\mathbf{V}_5}  \frac{\mu}{2}\Vert\mathbf{U}^{(k)}-\mathbf{V}_5-\mathbf{D}_5^{(k)}\Vert^2_F + \iota_{\mathcal{S}}(\mathbf{V}_5).
\end{equation*}
which can be conducted as follows
\begin{equation*}
	\mathbf{V}^{(k+1)}_5 = \left(\mathbf{U}^{(k)}-\mathbf{D}_5^{(k)}\right)+\mathbf{R}.
\end{equation*}
with $\mathbf{R}=\frac{1}{M}\left[\boldsymbol{1}_N^T-\boldsymbol{1}_M^T\left(\mathbf{U}^{(k)}-\mathbf{D}_5^{(k)}\right)\right]\otimes\boldsymbol{1}_M$ where $\otimes$ is a Kronecker product.

Finally the Lagrange multipliers $\mathbf{D}_1,\mathbf{D}_2,\mathbf{D}_3,\mathbf{D}_4,\mathbf{D}_5$ are updated using the following rules
\begin{equation*}
	\begin{aligned}
	& \mathbf{D}^{(k+1)}_1 = \mathbf{D}^{(k)}_1-\mathbf{E}\mathbf{U}^{(k+1)}+\mathbf{V}^{(k+1)}_1\\
    & \mathbf{D}^{(k+1)}_2 = \mathbf{D}^{(k)}_2-\mathbf{U}^{(k+1)}+\mathbf{V}^{(k+1)}_2\\
    & \mathbf{D}^{(k+1)}_3 = \mathbf{D}^{(k)}_3-\mathbf{V}^{(k+1)}_2\mathbf{W}+\mathbf{V}^{(k+1)}_3\\
    & \mathbf{D}^{(k+1)}_4 = \mathbf{D}^{(k)}_4-\mathbf{U}^{(k+1)}+\mathbf{V}^{(k+1)}_4\\
    & \mathbf{D}^{(k+1)}_5 = \mathbf{D}^{(k)}_5-\mathbf{U}^{(k+1)}+\mathbf{V}^{(k+1)}_5.
    \end{aligned}
\end{equation*}

\ifCLASSOPTIONcaptionsoff
  \newpage
\fi

\bibliographystyle{IEEEtran}
\bibliography{strings_all_ref,ref_all}

\end{document}